\documentclass[preprint2]{aastex}

\usepackage{graphicx}
\usepackage{hyperref}
\usepackage[english]{babel}
\usepackage{rotating}
\usepackage{ngerman}

\shorttitle{The collisional evolution of undifferentiated asteroids}
\shortauthors{Beitz et al.}
\slugcomment{Submitted to The Astrophysical Journal}

\begin{document}

\title{The collisional evolution of undifferentiated asteroids and the formation of chondritic meteoroids}
\author{E. Beitz, J. Blum}
\affil{Institut f\"ur Geophysik und extraterrestrische Physik, Technische Universit\"ut Braunschweig,\\Mendelssohnstr. 3, D-38106 Braunschweig, Germany}
\email{e.beitz@tu-braunschweig.de}

\author{M. G. Parisi}
\affil{Instituto Argentino de Radioastronom\'{\i}a (IAR), CCT-La Plata,	CONICET, CC N$^{o}$ 5 (1894), Villa Elisa, Argentina}
\affil{Facultad de Ciencias Astron\'omicas y Geof\'{\i}sicas (FCAG),	Universidad Nacional de La Plata (UNLP), Argentina}

\author{J. M. Trigo-Rodr\'{i}guez}

\affil{Meteorites, Minor Bodies and Planetary Science Group, Institute of Space
Sciences (CSIC-IEEC), Campus UAB Bellaterra, c/Can Magrans s/n, 08193
Cerdanyola del Vall\`{e}s (Barcelona), Spain}

\begin{abstract}

Most meteorites are fragments from recent collisions experienced in the asteroid belt. In such a hyper-velocity collision, the smaller collision partner is destroyed, whereas a crater on the asteroid is formed or it is entirely disrupted, too. The present size distribution of the asteroid belt suggests that an asteroid with 100 km radius is encountered $10^{14}$ times during the lifetime of the Solar System by objects larger than 10 cm in radius; the formed craters cover the surface of the asteroid about 100 times. We present a Monte Carlo code that takes into account the statistical bombardment of individual infinitesimally small surface elements, the subsequent compaction of the underlying material, the formation of a crater and a regolith layer. For the entire asteroid, 10,000 individual surface elements are calculated.  We compare the ejected material from the calculated craters with the shock stage of meteorites with low petrologic type and find that these most likely stem from smaller parent bodies that do not possess a significant regolith layer. For larger objects, which accrete a regolith layer, a prediction of the thickness depending on the largest visible crater can be made.  Additionally, we compare the crater distribution of an object initially 100 km in radius with the shape model of the asteroid (21) Lutetia, assuming it to be initially formed spherical with a radius that is equal to its longest present ellipsoid length. Here, we find the shapes of both objects to show resemblance to each other.

\end{abstract}



\section{Introduction}\label{sec:intro}
The meteorites are the best and largest source of available material as a means to study the formation and evolution of our Solar System, but remain to be allocated in their parent body context. The most primitive meteorites coming from undifferentiated asteroids, are called chondrites. However, even the most primitive among the meteorites are ejecta from recent collisions in the asteroid belt. This can be shown by comparing the absolute age of the chondritic material, which resembles best the age of the Solar System itself \citep{baker2005early}, and the cosmic ray exposure age (CREA) \citep{eugster2006irradiation}, which is a measure for the duration of the journey from the asteroid belt to the Earth. \citet{herzog2010cosmic} measured this time to be on the order of a few 10 Myrs, which is much shorter than the time the present meteorites have spent as part of their respective parent bodies. This short CREA is in agreement with the present flux of meteorites produced by objects emerging from the asteroid belt  via resonant phenomena (see, e.g., \citet{morbidelli2002origin}; \citet{morbidelli1998orbital}; \citet{Elia_Brunini2007}).  Bodies in the asteroid belt may enter the $\nu_6$, 3:1 and 5:2 resonances and stay there for a few Myrs to be then transferred to the Earth. Evidence pointing to these dynamic delivery mechanisms comes from the meteoroids with well-determined heliocentric orbit updated in \citet{trigo2015orbit}.

When a hyper-velocity impact occurs on an asteroid, a pressure wave is generated, which compacts the ejected material as well as the material beneath the formed crater. The large number of collisions over the lifetime of the Solar System leads to a continuous flux of Earth-crossing meteoroids and reduces successively the primitivism of the remaining asteroids. The collision-induced compaction of the bodies of the asteroid belt must be considered when comparing the properties of asteroids with those of meteorites. \citet{davison2013} showed that most of the mass of the asteroid belt disappeared after only 100 Myrs and that the size distribution of the present asteroid belt has not significantly changed from that time and can therefore be considered constant for most of the lifetime of the Solar System. Taking the size distribution of the present asteroid belt \citep{Elia_Brunini2007}, one can show that for instance an asteroid of 100 km in radius has been bombarded more than $10^{14}$ times by fragments with radii between 0.1 m and 22 km during the last 4.5 Gyrs, with the smaller impactors being much more numerous than the large ones.
\citet{o1985impact} studied the velocity distribution of impact ejecta and compared them with the escape velocity of parent bodies with different sizes. The authors found that 99.9 percent of the crater material that is produced by an impact at 5 {$\mathrm{km\,s^{-1}\,}$} on a asteroid with 100 km in radius is gravitationally recaptured. Following this, a regolith layer should be present on all larger objects in the asteroid belt down to sizes of a few km. Such small bodies only recapture about 50 percent of the crater ejecta. The packing density of the regolith layer was recently studied by \citet{schraepler2015} and is mostly independent of the size of the parent body and for not too small grains independent of the regolith grain size. Typical volume filling factors are $\phi=0.6$, close to the limit of random close packing. Here, the volume filling factor $\phi$ describes the ratio between the volume filled with material and the total volume of the body.
	
\citet{holsapple1993scaling} studied the formation of craters in hyper-velocity impacts and found that the crater size can reach up to ten times the size of the impactor, and mainly depends on the collision velocity and the physical properties of the target. When the number of collisions and the size and velocity distributions of the impactors are known, the crater coverage of the target asteroid can be computed. An asteroid with 100 km radius is covered $\sim 100$ times with craters of sizes $\sim 1$~m and above. Thus, the most recent collisions by which the present-day meteorites are produced cannot be considered to originate from {\em primitive} asteroidal material but stems from pre-compacted matter \citep{BlumEtal:2006}. In laboratory impact experiments, \citet{BeitzEtal2013} found that the shock wave, which originates from the impact, not only consolidates the target within the crater volume, but penetrates significantly deeper into the target below the crater bed where it leads to a consolidation of the remaining body. This compaction process changes the initial condition of all further impacts. Details on the experiments by \citet{BeitzEtal2013} will be given in Section \ref{sec:contr_from_exp_stu}.

The idea of this study is to predict the collisional evolution of a chondritic parent body undergoing a sequence of impacts over the lifetime of the Solar System. In Section \ref{sec:compaction_model}, we elaborate on our compaction model for hyper-velocity impacts. Section \ref{sec:collison_frequency} introduces the collision frequency of an asteroid of 100 km radius with bodies in the radius range between 0.1 m and 22 km. In Section \ref{sec:1d_vs_3d} we compare a 3-dimensional approach of surface cratering with a new 1-dimensional approach. In Section \ref{sec:Monte_Carlo_Code} we present our Monte-Carlo code to study the collisional evolution of this asteroid under bombardment, Section \ref{sec:results} presents the results of our study, in Section \ref{sec:discussion} we discuss these results in the context of meteoritic and asteroidal properties, and we conclude our findings in Section \ref{sec:conclusions}.

\section{The impact compaction model}\label{sec:compaction_model}

In this study, we present an evolution model for asteroids, which is mainly based on the results of laboratory compaction experiments. The asteroid belt is a dense and dynamical region of our Solar System. Due to intense resonances with Jupiter, the formation of a planet-sized rocky body beyond Mars was prevented, because destructive collisions were frequent among the objects of the asteroid belt, and also gravitational scattering induced by giant planets played a role. The present and the initial population of the asteroid belt \citep{Elia_Brunini2007} show a strong depletion of objects with radii smaller than 50 km. Thus, we expect any larger asteroids to mostly survive the bombardment with smaller objects for a duration 4.5 Gyrs. The by far largest number of collision partners is much smaller than the target size and, thus, intrinsically harmless for the survival of the asteroids.

In the first part of this Section, we will show that the hydrostatic pressure of asteroids up to 100 km radius is too low to compress the material to volume filling factors measured in chondrites ($\phi = 0.6-1$, see \citet{Macke2011}). In the second part of this Section, we recapitulate the results and constraints from the compaction experiments performed by \citet{BeitzEtal2013} and show how we apply them to the numerical simulation in this work. The second important part in this study is the cratering process in high-velocity impacts. The used equations and assumptions will also be presented there.

To take the effect of previous impacts on the outcome of present collisions into account, we assume an initially large (100 km radius) and porous ($\phi = 0.6$) parent body as predicted by \citet{morbidelli2009asteroids}. However, our results are easily scalable to other parent-body sizes.

\subsection{\label{sec:hydrostatic}The unimportance of hydrostatic compaction}

The assumption that the parent bodies are initially formed porous is supported by several different facts. \citet{britt1987asteroid} estimated the macro-porosity of asteroids by scaling the size of the measured asteroids with the bulk porosity of corresponding meteorites. The authors showed that the asteroids cover a wide range of volume filling factors but cluster for objects of about 100 km in size at volume filling factors of $\phi  \sim0.7$. The presence of foliation, which is frequently found in chondrites, originates from hyper-velocity impacts on {\em porous} material \citep{Gattacceca2005}.

Large celestial objects are subject to hydrostatic compression in their interiors that lead to a compaction of material and, thus, to a density increase towards the center. The hydrostatic pressure in the center of a body with radius $R$ and mass $M$ is given by $p_{\mathrm{c}}=\frac{3}{8 \pi} \frac{G M^2}{R^4}$, with $G$ being the gravitational constant. If the compressional strength of the material, $p_{\mathrm{comp}}$, is known, then one can calculate the maximum radius for which the material does not yield the hydrostatic pressure and gets
\begin{equation}\label{hydpr}
	R_{\mathrm{max}} = \sqrt{\frac{3 p_{\mathrm{comp}}}{2 \pi G \varrho^2}},
\end{equation}
with $\rho$ being the constant mass density in the interior of the body. From Fig. 11 in \citet{BeitzEtal2013}, we derive that the compressional strength of dusty material consisting of micrometer-sized monomer grains  is on the order of $p_{\mathrm{comp}} = 10^7$ Pa, i.e. hydrostatic or impact pressures exceeding this value are required to remove the microporosity within the dusty material. The mass density is in the range $\varrho \approx
1,000-2,000~\mathrm{kg~m^{-3}}$ for volume filling factors between $\phi = 0.3$ and $\phi = 0.7$. This leads to maximum radii in the range of 134 km - 268 km. This means that bodies with radii below $\sim 100$ km will not experience significant hydrostatic compression and will, thus, be able to sustain a constant mass density.

This is also true if the bodies were formed by gravitational collapse, following the model by \citet{2007Natur.448.1022J}. Here, planetesimals form by the gravitational collapse of pebble-sized dust aggregates, which undergo low-velocity mutual collisions during the contraction phase. \citet{joansson2014} calculated that due to frequent low-velocity collisions among the pebbles, the average collision speed stays below $1~\mathrm{m~s^{-1}}$, which corresponds to dynamical pressures on the order of $1$ kPa, well below the material strength of the pebbles (see above). However, pressures as low as $\sim 10^5$ Pa are sufficient to deform the dust aggregates. The central pressure inside a body with 100 km radius is typically few MPa so that the porosity on size scales comparable to the dust-aggregate sizes is removed (see Fig. 11 in \citet{BeitzEtal2013}). Thus, for pressures above $\sim 10^5$ Pa, the volume filling factor is expected to be $\phi \approx 0.6$, throughout most of the volume of the body. Mind that the removal of the remaining microporosity is only possible for pressure levels of $p \geq p_{\mathrm{comp}} = 10^7$. It is therefore appropriate to assume that the initial asteroid-sized planetesimals were spatially homogeneous in density and porous, which is what we will do in our model as described below.

\subsection{Constraints from experimental studies}\label{sec:contr_from_exp_stu}

\begin{figure}[htp]
	\begin{center}
		 \includegraphics[width=8cm]{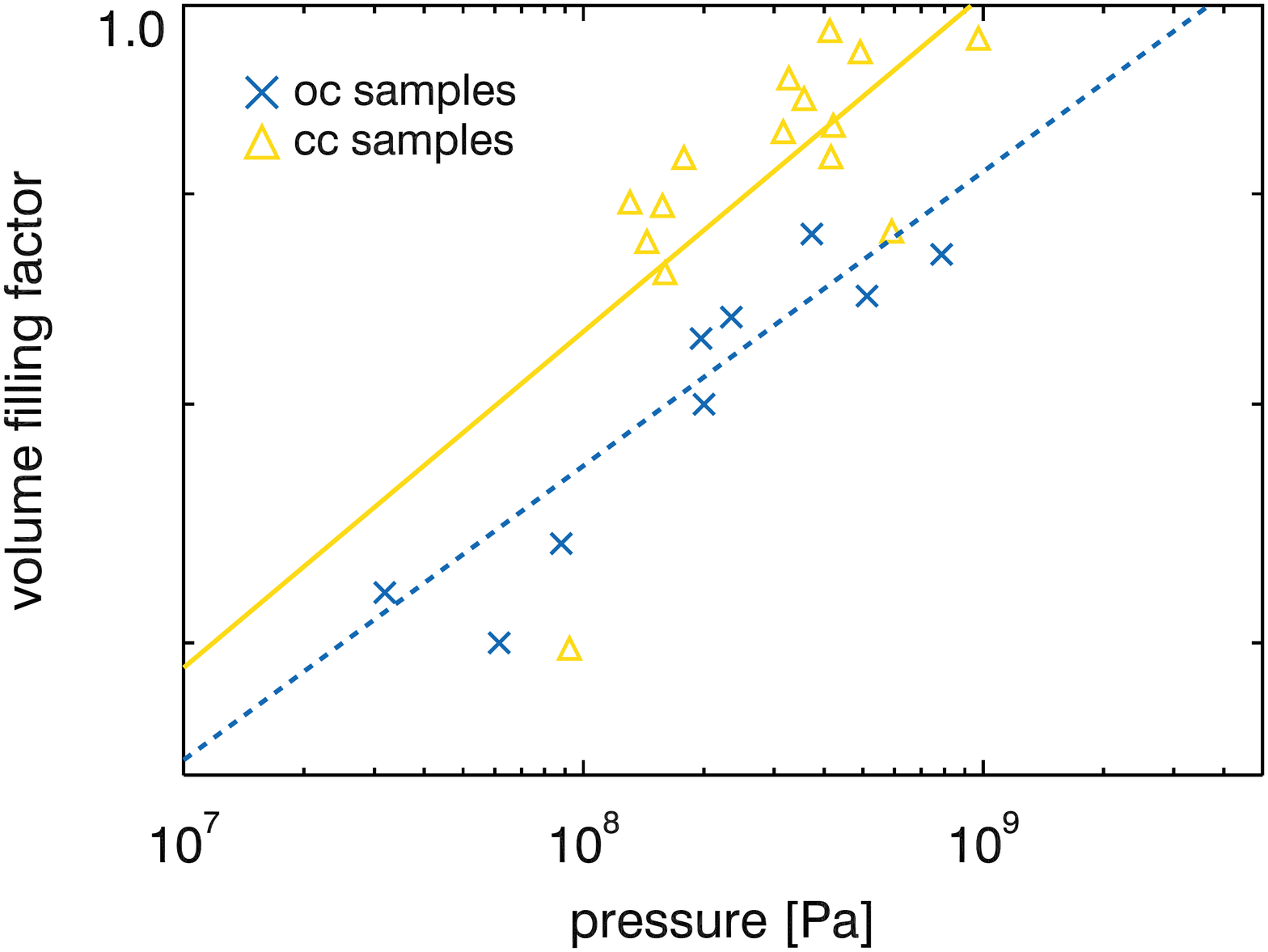}
		\caption{\label{fig:compression_chondrites_monte} Relation between the volume filling factor and the maximum pressure for dust-dominated and chondrule-dominated samples, using the modified data of \citet{BeitzEtal2013}. The yellow curve is fitted to the dusty carbonaceous chondrite (CC) analog samples shown as triangles and has a slope of 0.082. The blue curve is fitted to the analogs of the ordinary chondrite (OC) samples and has a slope of 0.072.}
	\end{center}
\end{figure}

The numerical model presented below is based on the impact experiments and constraints from our previous work \citep{BeitzEtal2013}, in which we studied the compaction of chondritic analog material in high-velocity impact experiments. The samples consisted of $\mu$m-sized silica grains, which can be seen as an analog material for the matrix of chondrites, and of mm-sized solid alumina and glass beads, which serve as analogs for the chondrules in chondrites. Thus, mixing the two components in different proportions, the compaction behavior of ordinary chondrites and carbonaceous chondrites was studied separately by using chondrule fractions of more than 80 percent or less than 50 percent, respectively. The chondritic material was filled into nylon tubes, which were enshrouded by massive steel housings to prevent the nylon tubes from breaking apart when being impacted by the projectiles. The projectiles were aluminum rods with varying lengths, whose diameter were kept constant and were only slightly smaller than the inside diameter of the nylon tubes. By choosing these parameters, the whole energy was dissipated in compacting the target and the formation of a crater was prevented. After the impact, the compacted target was analyzed using computer-aided X-ray tomography. The degree of compaction was analyzed as a function of depth and we found the highest consolidation close to the point of impact. To calculate the impact pressure, the impedance-matching method from \citet{Melosh1989} was adopted. This theory was extended to be porosity-dependent and provides the maximum impact pressure at the projectile-target interface, i.e.
\begin{equation}
	p_\mathrm{max} =p(\phi)_\mathrm{p/t} = \varrho_\mathrm{p/t} \cdot
	(C(\phi)_\mathrm{p/t} + S_\mathrm{p/t} \cdot u(\phi)_\mathrm{p/t}) \cdot
	u(\phi)_\mathrm{p/t}. \label{eq:pressure_melosh}
\end{equation}
Here, $u_\mathrm{p/t}$ is the ``particle velocity'' introduced by \citet{Melosh1989},  $S_\mathrm{p/t}$ denotes a material constant related to the Gr\"uneisen parameter, $\varrho_\mathrm{p/t}$ is the mass density of the projectile and target, and $C(\phi)_\mathrm{p/t}$ is the porosity-dependent sound speed (see details in \citet{BeitzEtal2013}). The indices p/t denote that the variables are to be taken for the projectile or the target, respectively. In Fig. \ref{fig:compression_chondrites_monte}, the correlation between this pressure and the volume filling factor is given as measured in the experiments for dust-dominated and chondrule-dominated samples, which can be regarded as analogs for carbonaceous and ordinary chondrules, respectively (see above). In the experiments, we found that the degree of compaction decreases with increasing depth within the sample and is a function of the length $h$ of the impacting projectile, following $p \varpropto h^{a}$. This is in good agreement with the findings of previous studies by \citet{nakazawa2002experimental}. In their experiments, the authors found the exponent $a$ for the relevant pressure range to be $a=1.8\pm0.2$.

Eq. \ref{eq:pressure_melosh} can be simplified if the projectile and target material and density are identical. It then reads
\begin{equation}
	p_\mathrm{max} = \varrho_\mathrm{p=t} \cdot (C(\phi) + S_\mathrm{p/t} \cdot v_\mathrm{imp}) \cdot 0.5\cdot v_\mathrm{imp} \label{eq:pressure_melosh_easy},
\end{equation}
using the impact velocity $v_\mathrm{imp}$ instead of the abstract ``particle velocity''. We use the simplified Eq. \ref{eq:pressure_melosh_easy} as an approximation to Eq. \ref{eq:pressure_melosh} under the assumption that all objects and, thus, collision partners in the asteroid belt are statistically exposed to the same number of collisions per unit area during their evolution, and are, thus, compacted in the same way. In this study we will focus on ordinary chondrite parent bodies.

\subsection{\label{crater}Cratering law}

In  the following, we briefly describe the cratering law that  is used to evaluate the production of craters on large asteroids by impacts with the population of small bodies in the asteroid belt during the age of the Solar System.

We follow the pi-group scaling of \cite{holsapple1993scaling} to determine the size of the crater. There are two radii in a transient crater that can be identified, $R_{tr}$ and $R_{tc}$, respectively. $D_{tr}=2 R_{tr}$ is the corresponding diameter of the transient crater measured from rim crest to rim crest, while $D_{tc}=2 R_{tc}$ is the diameter of the transient crater measured at the pre-impact surface. It was found that $D_{tr}=1.3 D_{tc}$ (\cite{collins2005}, \cite{holsapple1993scaling}). \cite{holsapple1993scaling} gives for $R_{tc}$

\begin{equation}
	\left(\frac{\rho_{t}}{m}\right)^{(\frac{1}{3})} R_{tc} =
	(K_1)^{(\frac{1}{3})}\left[\pi_2+
	 \bar{\pi_3}^{\frac{(2+\mu)}{2}}\right]^{\frac{-\mu}{(2+\mu)}}.
	\label{eq:r_tctr}
\end{equation}

Here, $\pi_2$ and $\pi_3$ are the ratio of the lithostatic pressure at a depth equal to one projectile radius to the initial dynamic impact pressure and the ratio between the crustal material strength and the initial dynamic impact pressure, $\rho_{t}$ and $m$ are the mass density of the target body and the impactor mass, and $K_1$ and $\mu$ are material-dependent fitting parameters, respectively.

Then, if $\pi_2> \bar{\pi_3}^{\frac{(2+\mu)}{2}}$, impacts are in the gravitational regime, while, if $\pi_2<\bar{\pi_3}^{\frac{(2+\mu)}{2}}$, impacts are in the strength regime.

The diameter of a transient crater decreases with increasing obliquity, all other factors remaining constant. Expressed in terms of crater volume $V$, it has been found that $V \propto \sin\theta$ (e.g. \cite{gault1978experimental}), where $\theta$ is the impact angle (i.e., $\theta=90\deg$ for vertical incidence). Then, taking into account the impact obliquity, the radius of the crater in the gravitational and    strength regimes may be obtained from Eq. \ref{eq:r_tctr}. In the gravitational regime, we neglect the second term in Eq. \ref{eq:r_tctr} and the radius of the transient crater is then given by

\begin{equation}
	R_{tcg}= C g^{-\left(\frac{\mu}{\mu+2}\right)} v_{imp}^{\left(\frac{2 \mu}{2+\mu}\right)}
	r_{imp}^{\left(\frac{2}{2+\mu}\right)} (\sin{\theta})^{\frac{1}{3}},
	\label{eq:rcgt}
\end{equation}

while in the strength regime, the first term in Eq. \ref{eq:r_tctr} is neglected to obtain the radius of the crater

\begin{equation}
	R_{tcs}= C \left(\frac{\bar{Y}}{\rho_{t}v_{imp}^{2}}\right)^{(\frac{-\mu}{2})}
	r_{imp}~(\sin{\theta})^{\frac{1}{3}}.
	\label{eq:rcst}
\end{equation}

Here, $v_{imp}$, $r_{imp}$, $g$ and $\bar{Y}$ are the impact velocity, the impactor radius, the surface gravity of the target and the effective material strength, respectively. Moreover, $C=  (K_1)^{(\frac{1}{3})} (\frac{4\pi}{3})^{(\frac{1}{3})}$. \cite{holsapple1993scaling} obtained $K_1=$ 0.24 for dry soil and 0.2 for soft rock, while \cite{schmidt1987some} give $K_1=$ 0.33 for dry soil. From these values of $K_1$, $C$ results as being on the order of unity for both materials so that we further assume $C = 1$. For $(\sin{\theta})^{1/3}$ we take the average value of $<(\sin{\theta})>^{1/3}= (\pi/4)^{1/3}$.

We carried out calculations for both materials, dry soil and soft rock. For dry soil, we assume $\mu$=0.41 and $\bar{Y}$=0.18 MPa \citep{holsapple1993scaling}. For soft rock, \cite{holsapple1993scaling} gives $\mu$=0.55 and $\bar{Y}$=7.6 MPa. We take $\mu$=0.5641 and $\bar{Y}$=7.6 MPa for soft rock, arriving at the expressions for $D_{tc}$ given by \cite{davison2013} and \cite{collins2005}  in the gravitational regime and by \cite{asphaug2008critical} in the gravitational and strength regimes. We assume that the transition  between the strength and the gravitational regimes occurs at $r_{imp}$= 165 m for dry soil and at $r_{imp}$=  724 m for soft rock.
The depth $d_{tc}$ and rim height $h_{tr}$ of the transient crater measured from the pre-impact surface are $d_{tc}= R_{tc}/\sqrt2$ and $h_{tr}=0.07R_{tr}$ \citep{collins2005,holsapple1993scaling}. For impactors in the strength regime, we assume that the final crater radius $r_c = R_{tcs}$, with $R_{tcs}$ given by Eq. \ref{eq:rcst} and the crater depth $h_c=d_{tc}$, i.e., $h_c=R_{tcs}/\sqrt2$. In the  gravitational regime, the transient crater is an intermediate step in  the development of the  final crater, which may be simple or complex.  The transition from simple to complex craters is known to occur at 3.2 km on Earth and at 18 km on the Moon \citep{davison2013,collins2005}.   The simple-to-complex transition diameter is given by \citep{davison2013}

\begin{equation}
	d_{sc} =\frac{g_{moon} \rho_{moon} d_{sc_{moon}}}{g \rho_t},
	\label{eq:com}
\end{equation}

where $g_{moon}$, $\rho_{moon}$, and $d_{sc_{moon}}$ are the surface gravity, density and simple-to-complex transition diameter on the Moon. From Eq. \ref{eq:com}, we obtain $d_{sc} \sim 1,000$ km for a target of radius 100 km. Then, all the final craters on our asteroidal target are simple craters and there are no complex craters. For simple craters, the collapse process is well understood, where highly brecciated and molten rocks that were originally pushed out of the opening crater slide back down the steep transient cavity walls, forming a melt-and-breccia lens at the base of the crater  \citep{collins2005}. To derive the final crater radius and depth for simple craters, we follow \cite{collins2005}. They obtained that the rim -to-rim diameter of a simple crater is $D_{fr} = 1.25 D_{tc}$. For  impactors in the gravitational regime, we then assume that the final crater radius is $r_c= 1.25 R_{tcg}$, with $R_{tcg}$ given by Eq. \ref{eq:rcgt}. The rim height, $h_{fr}$, above the pre-impact surface and the unbulked breccia lens volume, $V_{br}$, were derived by \cite{collins2005} to be
\begin{equation}
	h_{fr}= 0.07 \frac{D_{tc}^{4}}{D_{fr}^{3}},
	\label{hfr}
\end{equation}
and
\begin{equation}
	V_{br}= 0.032 D_{fr}^{3},
	\label{vbr}
\end{equation}
respectively.

Then, the breccia lens thickness $t_{br}$ may be expressed in the following form \citep{collins2005}

\begin{equation}
	t_{br}= 2.8 V_{br} \left(\frac{d_{tc}+h_{fr}}{d_{tc}D_{fr}^{2}}\right).
	\label{tbr}
\end{equation}

For impactors in the gravity regime, we assume that the final crater depth $h_c=d_{fr}$, where the crater depth $d_{fr}$ is measured from the crater floor (above the breccia lens) to the rim crest and is given by

\begin{equation}
	d_{fr}= d_{tc}+h_{fr}-t_{br}.
	\label{tbr2}
\end{equation}

\section{Collision rate and impact velocity}\label{sec:collison_frequency}

In this section, we will describe how the collision frequency for an asteroid with primordial diameters between 70 and 268 - 536 km is calculated. The lower bound of 70 km in the asteroid diameter is taken as the smallest size without likely catastrophic fragmentation during the age of the Solar System. A catastrophic collision of a target body occurs when it collides with an impactor that carries sufficient energy to cause the largest remnant to possess 50\% of the initial target mass. \cite{jutzietal2010} calculated that a catastrophic collision occurs when a projectile with $\sim 27$ km radius hits a porous target with 100 km radius. For target radii of 10 km, an object of $\sim 1$ km radius is sufficient to catastrophically disrupt the target. The present average time between catastrophic collisions for asteroids with diameters between 1 m and 10,000 m is in the range between 3-6 Myrs and 1.7-2.7 Gyrs  \citep{Elia_Brunini2007}.For objects with radii of $\sim 35$ km, the mean time between catastrophic collisions 3.8-4.6 Gyrs, i.e., comparable to the age of the Solar System. Thus, our treatment is valid for targets of this minimum radius of $\sim$ 35 km. The upper bound of 268 - 536 km in diameter is given by the condition of hydrostatic compaction (see Section \ref{sec:hydrostatic}). Taking into account these bounds, we henceforth assume a target asteroid of 100 km radius, but our results can easily be applied to other asteroid sizes in the range given above. We also assume henceforth that the asteroid was small enough or formed late enough to escape melting and, thus, differentiating \citep{2003Natur.422..154Y}.

We study the collision frequency and the size frequency distribution of the projectiles (small asteroids) impacting the target (large asteroid) during the age of the Solar System. The collision frequency of those impactors can be derived from the smoothed  number-frequency distribution of the present asteroid belt \citep{Elia_Brunini2007}.

The impact rate is calculated assuming two constant values for the impact velocity $v_{imp}$, namely 3 $\rm km s^{-1}$ and 5 $\rm km s^{-1}$, respectively, which are below and above the mean collision velocity of the asteroid belt (see Fig. \ref{fig:vel_dist}). These limits simulate a conservative and more progressive number of collisions over the lifetime of the Solar System. It should be mentioned that the assumption of constant impact velocity is only used to derive the impact rate (see Eqs. \ref{eq:imp} and \ref{eq:vol}) and not for the crater properties.

\citet{davison2013} obtained that most of the mass of the asteroid belt disappeared after only 100 Myrs and that the size distribution of the present asteroid belt can be considered as constant during the age of the Solar System. This is also in agreement with the results obtained by  \citet{bottke2005fossilized}, which indicate that the main-belt size distribution is predominately a fossil from the early stages of the Solar System. Thus, we take the ``Final Main Belt population'' of Fig. 7a of \citet{Elia_Brunini2007} as the size frequency distribution of the projectiles of radius $r$ impacting a target of radius 100 km during the age of the Solar System.

The number of impacts on the target was computed from the data of that Figure, which were kindly provided to us by de El\'{\i}a (pers. comm.). These data comprise the total number of asteroids $N(r_i,r_{i+1})$ in a bin of impactor radius $[r_i, r_{i+1}]$ in the main belt, extending from 2 AU (approximate location of the $\nu_6$ secular resonance) to 3.27 AU (location of the 2:1 mean motion resonance). In addition, we fit from these data a differential power-law size-frequency distribution of index $p \sim -2.8$ for 0.1 m $\le r \le$ 135 m and $p \sim -1.7$ for 135 m $\le r \le$ 22 km. The computations by \citet{Elia_Brunini2007} take into account the action of the Yarkovsky effect, the effect of the ``powerful resonances'' $\nu_6$, 3:1, 5:2, and 2:1, and collisional fragmentation. We cut the size frequency distribution at a minimum projectile radius of 0.1 m, since smaller bodies are strongly affected by Poynting-Robertson drag, considerably reducing their lifetimes in the asteroid belt. We assume a maximum projectile radius of 21,976 m, since collisions with larger objects would lead to a crater depth larger than the 100 km radius of the assumed asteroid. This upper impactor size limit is in full agreement with the collision probability calculations by \citet{davison2013} (their Fig. 4), who predict only a probability of $\sim 9\%$ for impacts with projectiles above 20 km radius. The projectiles in the size range 0.13264 m $\le r \le$ 21,976 m are distributed in 53 logarithmically equidistant size bins following $r_{i+1} = 1.26~r_{i}$ for $i = 1 \ldots 52$ \citep{Elia_Brunini2007}.

\begin{figure}[htp]
	\begin{center}
		\includegraphics[width=8cm]{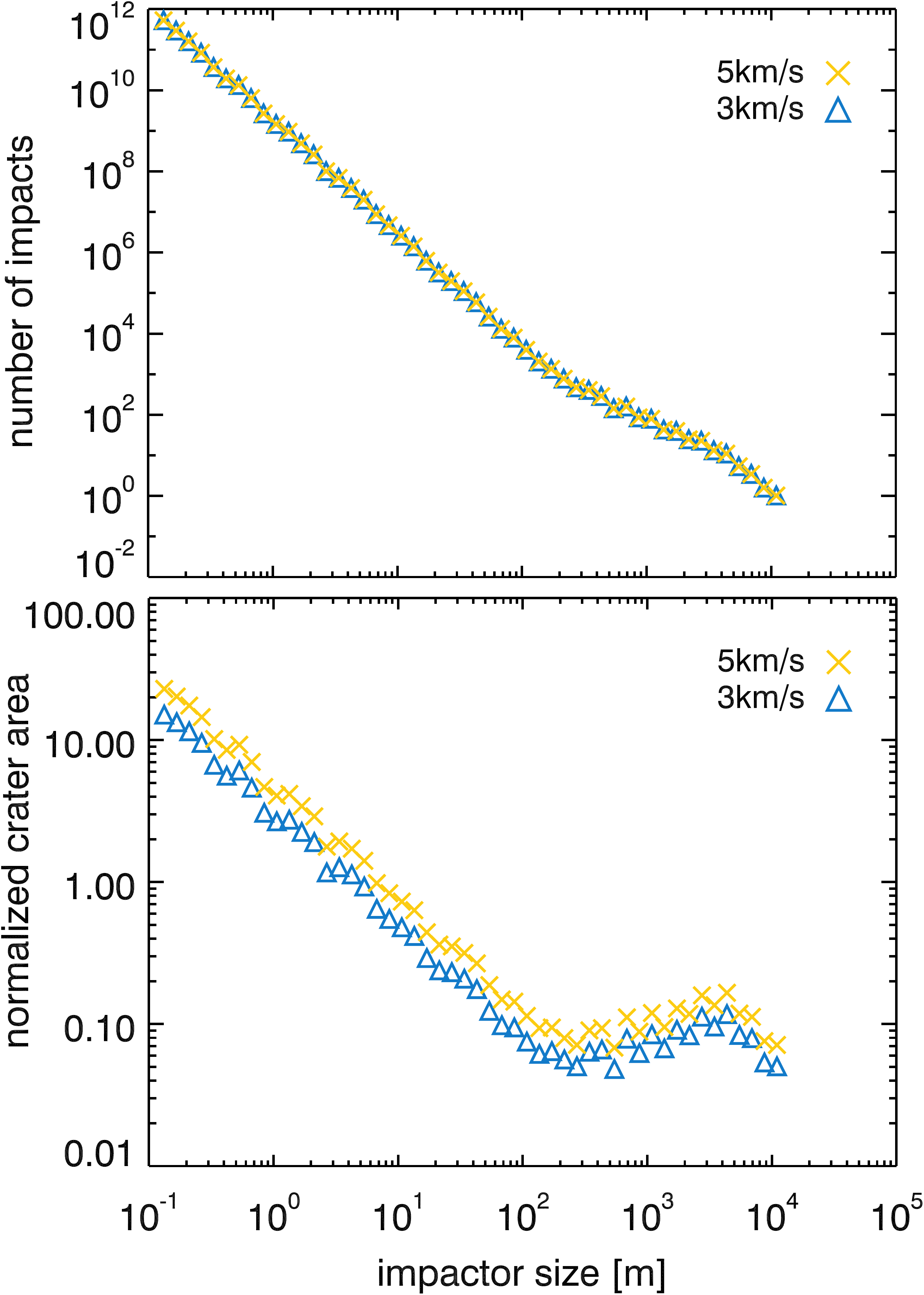}
		\caption{\label{fig:coll_number} Top: Differential number of impacts of impactors with radii in the range between 0.1 m and 22,000 m on a 100 km radius parent body over 4.5 Gyrs at a constant impact velocity of 3 and 5 $\rm km s^{-1}$, using the material properties for dry soil and soft rock given by \citet{holsapple1993scaling}. Bottom: Differential coverage of the target surface by craters resulting from the impactors shown at the top. The total coverage of the asteroid surface by craters of all sizes is 85, 152 and 187 times, for soft rock at 3 and 5 $\rm km s^{-1}$ and dry soil at 5 $\rm km s^{-1}$. The behavior for dry soil is only shown for comparison. The bin size in both cases is a factor 1.26.}
	\end{center}
\end{figure}

\begin{figure}[htp]
	\begin{center}
		\includegraphics[width=8cm]{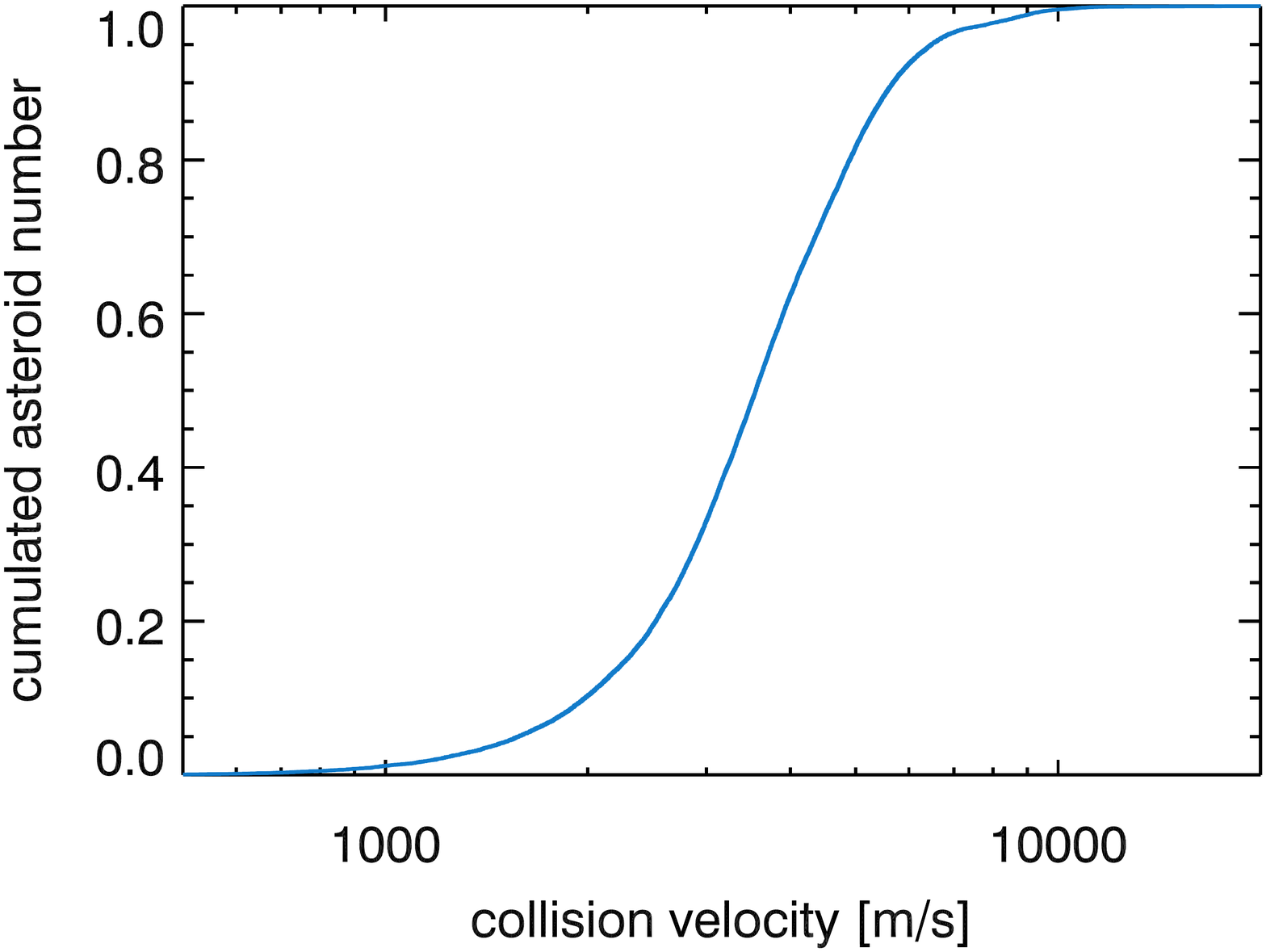}
		\caption{\label{fig:vel_dist} Velocity-frequency distribution of the present asteroid belt calculated as the dispersion velocity from the distribution of about half a million objects in the asteroid belt with known orbital elements and semi axis between 2.2 and 3.75 AU.}
	\end{center}
\end{figure}

The number of impacts of projectiles in the radius range $(r_i,r_{i+1})$ on a target of radius $R$ per unit time is given by the flux of projectiles onto the target times its collision cross section. The former is the product of the number density of the impactors, $N(r_i,r_{i+1})/V$, and the impact speed, $v_{\rm imp}$, with which the projectiles hit the target. The latter is given by $\pi R^{2}$. Thus, we get the number of impacts of projectiles in the radius range $(r_i,r_{i+1})$ per unit time
\begin{equation}
	\frac{{\rm d}N_{\rm p}(r_i,r_{i+1})}{{\rm d}t}=\frac{N(r_i,r_{i+1}) \pi R^{2} v_{\rm imp}}{V},
	\label{eq:imp}
\end{equation}
\noindent where the unit volume $V$ is given by
\begin{equation}
	V=4 \pi a_o \Delta{a_o} H,
	\label{eq:vol}
\end{equation}
and $a_o = 2.635$ AU, $\Delta{a_o} = 0.635$ and $H = 3.7\cdot10^7$ km or $H = 6.2\cdot10^7$ km being the mean semimajor axis of the asteroid belt, its half width and its mean thicknesses using mean velocities of 3 and 5 $\rm km\,s^{-1}$, respectively (see above). As shown in Fig. \ref{fig:coll_number}, the total number of impacts is almost independent of the assumed impact velocity, the latter influencing the coverage of the targets surface only. It should be mentioned that the impact-velocity dependence of the impact rate shown in Eq. \ref{eq:imp} is caused by the excitation of the orbits, which results in enhanced eccentricities and inclinations and, thus, in higher impact speeds (see Eq. \ref{eq:vei}).

The total number of impacts in each impactor size bin over the age of the Solar System is computed as
\begin{equation}
	N_{\rm col}(r_i,r_{i+1})=\frac{{\rm d}N_p(r_i,r_{i+1})}{{\rm d}t} 4.5 \rm Gyrs.
	\label{eq:ntot}
\end{equation}
The number of ``complete surface coverages'' by craters resulting from those impacts is
\begin{equation}
	N_{\rm s}(r_i,r_{i+1})= \frac{r_{\rm c}^{2} N_{\rm col}(r_i,r_{i+1})}{4 R^{2}},
	\label{eq:nshell}
\end{equation}
\noindent where $r_{\rm c}$ is the crater radius caused by projectiles in the size range $(r_i,r_{i+1})$, given by Eq. \ref{eq:rcgt} or Eq. \ref{eq:rcst}.
In the top panel of Fig. \ref{fig:coll_number}, the number of collisions $N_{\rm col,3}$ and $N_{\rm col,5}$ for two constant collision velocities of 3 and 5 $\rm km\,s^{-1}$ are shown. The lower panel of Fig. \ref{fig:coll_number} denotes the number of ``complete surface coverages'', which describes how many times on average each surface element is statistically part of a crater and, thus, affected by an impact for a given projectile size. Due to the dependence of the crater size on the impact velocity (see Sect. \ref{crater}), the surface coverage is higher for impacts occurring at 5 $\rm km\,s^{-1}$. The total surface coverage for projectiles of all radii  between 0.1 m and 22,000 m is 152 and 187 times using the material properties for dry soil and soft rock given by \citet{holsapple1993scaling} and a mean impact velocities 5 $\rm km\,s^{-1}$. Reducing the mean impact velocity to 3 $\rm km\,s^{-1}$, the total surface coverage drops to 85 times using soft rock.

To derive the number of impacts and crater sizes on the target asteroid, the collision velocity $v_{\rm imp} = \sqrt{(v_{\rm e}^{2}+v_{\rm inf}^{2})}$ needs to be derived, where $v_{\rm e}$ and $v_{\rm inf}$ are the escape velocity from the target body and the relative velocity between projectile and target at large distances and averaged over an epicycle and over a vertical oscillation, respectively. For an asteroid of 100 km radius, the escape velocity is on the order of 0.2 $\rm km\,s^{-1}$ and, thus, small compared to the typical collision velocities in the asteroid belt. Therefore, it was neglected in the subsequent calculations. Thus,

\begin{equation}
    v_{\rm imp} \approx v_{\rm inf} =\sqrt{\varepsilon^2 + i^2} \cdot v_{\mathrm{kep}},
    \label{eq:vei}
\end{equation}

with $\varepsilon$, $i$ and $v_{\mathrm{kep}}$ being the mean orbital eccentricity of the colliding objects, the inclination (assumed to be small so that $\sin i \approx i$) and the Keplerian velocity, respectively. The Keplerian velocity is calculated at the mean semimajor axis of $a_o = 2.735$ AU, i.e., for an extent of the asteroid belt between 2.2 and 3.27 AU. Assuming that the local dispersion velocity with respect to the local Keplerian speed is isotropic ($v_{\rm inf,x} = v_{\rm inf,y} = v_{\rm inf,z}$), we get $\varepsilon^{2} v_{\rm kep}^{2}=v_{\rm inf,x}^{2}+v_{\rm inf,y}^{2}$ and $i^{2} v_{\rm kep}^{2}=v_{\rm inf,z}^{2}$, $\varepsilon= 2 i$ \citep{Parisi2013} and the height of the disk $H= a_o i$ with $i= v_{inf}/(\sqrt{6} v_{kep})$. To determine the distribution of the collision velocities of the asteroid belt, we used the orbital parameters of $\sim 500,000$ asteroids, provided by \hyperlink{http://ssd.jpl.nasa.gov/dat/ELEMENTS.NUMBR.}{http://ssd.jpl.nasa.gov/dat/ELEMENTS.NUMBR.} and shown in Fig. \ref{fig:vel_dist}.

\section{The Monte Carlo code}\label{sec:Monte_Carlo_Code}
In this section, the simulation code is presented. To describe the impact history on the surface and in the interior of an asteroid, we randomly pick an infinitesimally small point on the asteroid. For this point, we calculate for short time intervals the probability to be located within a newly formed crater. Depending on the crater size and the particular location of our arbitrary surface point within the crater, we calculate the excavated layer thickness and pressure decrease with increasing depth under the crater floor. The material below the crater floor is then compacted according to our experimental findings presented in \citet{BeitzEtal2013}. This procedure is repeated until 4.5 Gyrs of simulation time have passed. As a particular surface point is on average only covered a few times by small craters and even less by large ones (see Fig. \ref{fig:coll_number} bottom), we repeat the procedure for up to 10,000 surface points to also statistically capture the large craters. Each of these 10,000 surface points is treated individually with its own list of impacts generated according the impact probability distribution. The only shortcoming from this parallel approach is that we do not allow two or more of the surface elements to be part of the same crater. In this section, we will describe our Monte Carlo code with which the outcome of the bombardment was investigated, the approximations made, and the tests performed.

We derived the probability that a surface element falls within the crater made by an impact of a projectile with a given size from the crater coverage shown in the lower part of Fig. \ref{fig:coll_number}. We divided the total lifetime of the asteroid of 4.5 Gyrs in $10^6$ time steps of 4,500 yrs each and approximated the probability for being hit by an impactor of a given size by the crater coverage divided by $10^6$. For a total crater coverage of $\sim 100$, this gives an average probability of $\sim 10^{-4}$ per time step, small enough to prevent double impacts.

The code is divided into two individual parts. In the first simulation part, a list with the collision sequence for each surface element is produced, consisting of the impactor size, the collision time, the position of the surface element within the crater, and the collision velocity. In the second part of the simulation, each surface element is exposed to its sequence of impacts according to the list generated before and the physical alterations within the 100 km long, one-dimensional volume element, i.e. compaction, release of crater material and regolith deposition (if applicable), are applied.

For the first part mentioned above, it is first determined whether or not an impact occurs in the given time step of 4,500 yrs. Starting with the largest size bin of impactors, we draw a random number from a uniform distribution and compare it to the probability of a collision as described above. If the random number is larger than the probability, the procedure is continued with the second largest impactor until either a collision is determined (random number smaller than probability) or the smallest size bin is reached. In the latter case, no impact happens during this time step. If an impact is detected, the time and impactor size are stored. In this case, another random number is generated from which the impact velocity is randomly selected from the distribution shown in Fig. \ref{fig:vel_dist} (see Sect. \ref{sec:collison_frequency}). To account for the reduction of the impact velocity due to non-normal impacts, the impact angle $\theta$ is randomly generated and the impact velocity is reduced by multiplying the impact velocity with $sin(\theta)$, following \citet{gault1978experimental}. Finally, a last random number, drawn from a squared distribution, decides where inside the crater the surface element is located. Here, a value of 0 denotes the crater center for which the crater depth is maximal, whereas a value of 1 represents the crater rim with no excavated material.

\begin{figure}[htp]
	\begin{center}
		\includegraphics[width=8cm]{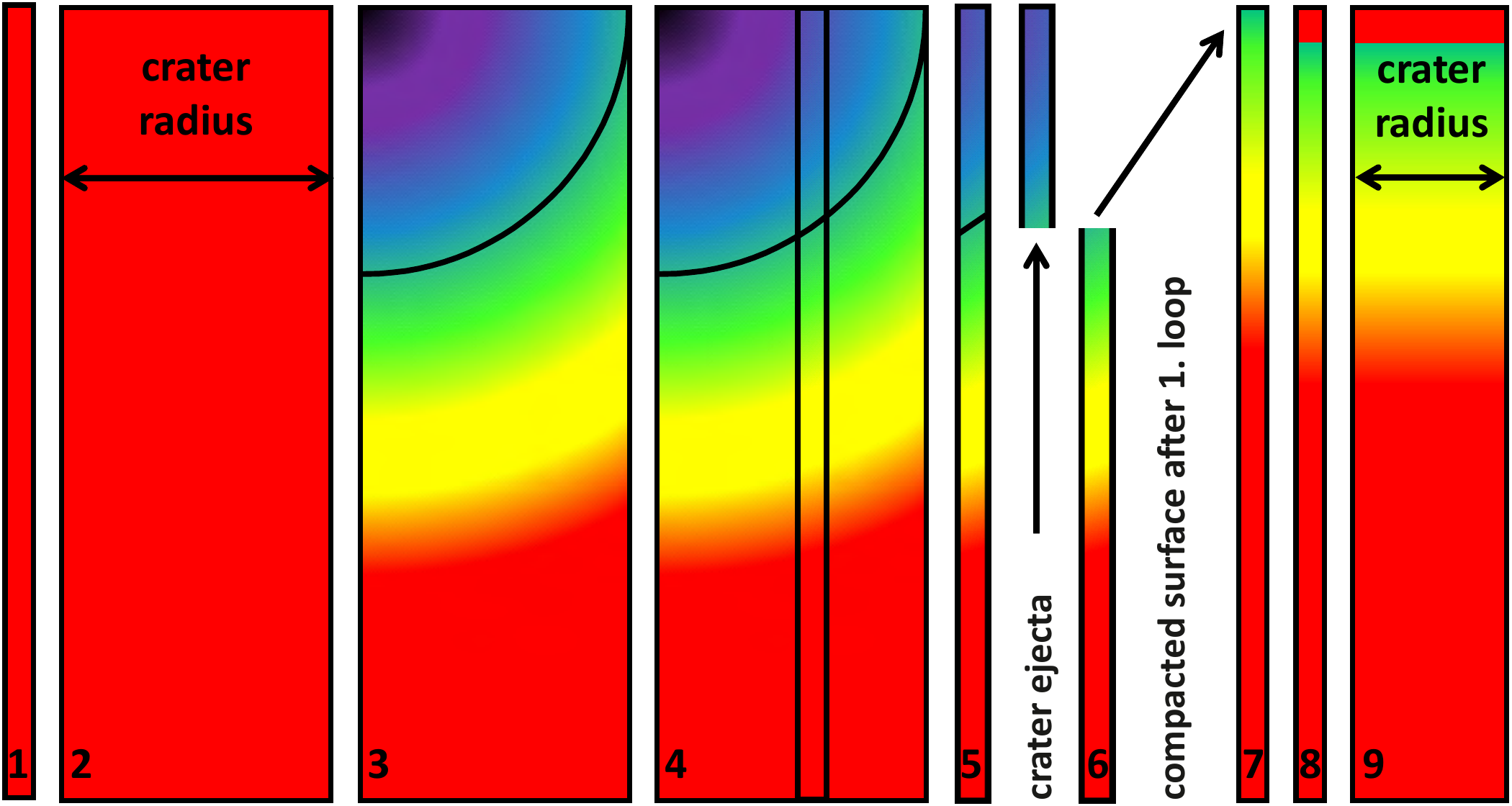}
		\caption{\label{fig:model} Schematics of the Monte Carlo simulations described in this paper. Steps 1-7 are performed sequentially for all collisions on this particular surface element, step 8 is only performed if regolith deposition is considered and step 9 is analog to step 2 for the subsequent impact. Colors denote the volume filling factor according to the color table used in Fig. \ref{fig:monte_movie}. 1: The initial surface element extending 100 km into the interior of the asteroid with an initially constant volume filling factor of 0.6. 2: For the selected impact, the surface element is horizontally blown up to the resulting crater radius while keeping its depth constant. 3: The impact pressure is calculated according to Eq. \ref{eq:pressure_melosh_easy} for the central impact point (upper left corner in the sketch); the pressure decreases with distance $h$ to the impact point according $p\varpropto h^{-2}$ in all directions until the corresponding volume filling factor reaches the background value. The solid curve denotes the crater bottom. 4: The position of the surface element in the crater is determined by a random number drawn from a squared distribution. 5: At the chosen position, a linear element is cut out of the two-dimensional cross-section array. 6: The linear element is cut in two pieces at the position of the local crater bottom (solid curve in 4, 5 and 6), with the upper part being the ejecta from the crater and the lower part being the compacted remaining material. 7: The remaining bottom part is filled up from below with uncompacted asteroidal material and denotes the initial element for the subsequent impact. 8: If the effect of re-accretion of ejecta material is studied, a layer of regolith is now added on top of the surface element. Its thickness is calculated by the mean amount of crater ejecta from all 10,000 elements for the time until the next impact. 9: Analog to step 2, the new element is extended to the crater radius of the following impact. Here, the volume filling factor distribution is assumed to be independent of the radial coordinate. }
	\end{center}
\end{figure}

In the second part of the simulation, the evolution for each of the $10^4$ surface elements is calculated individually, following the collision-sequence list produced in the first part. For each impact of the list, we calculate the impact pressure, the propagation of the shock wave with the resulting change in volume filling factor, as well as the loss of material due to the cratering process. This is performed for each collisions following the steps 1-7, which are illustrated in Fig. \ref{fig:model}. Step 8 is only applied when the re-accretion of regolith is considered. Step 9 is analog to step 2 for the subsequent impact. In the following, we describe the individual steps of the simulation. 1: Before the impact, the considered element is described by a one-dimensional array with $10^5$ 1-m deep entries, corresponding to a total depth of $10^5$ m with an initial volume filling factor. 2: For the considered impact, the element is extended to a two-dimensional array, with the additional radial dimension given by the crater radius calculated for the impact condition. This is justified as most craters on asteroids are extremely circular. 3: The impact pressure is calculated according to Eq. \ref{eq:pressure_melosh_easy} with the parameters $\varrho_\mathrm{p=t}=3,000\mathrm{kg/m^3}$,  $S_\mathrm{p/t}=1$,	 $C(\phi) = 7.0\cdot \exp\left(6.5 \cdot \phi\right)$  and the impact velocity taken from the list of collisions derived in the first part of the simulation. To calculate the resulting distribution of the volume filling factor from the impact-pressure distribution, the power law which was experimentally determined by  \citet{BeitzEtal2013} for ordinary chondrites is used (see Fig. \ref{fig:compression_chondrites_monte}). The calculated maximum pressure is applied to the upper left corner (central impact point) of the two-dimensional array and the radially symmetric decrease of the pressure is calculated by $p\varpropto h^{-2}$, with $h$ being the distance from the impact point, until the resulting volume filling factor falls below the value of the pre-impact situation. We approximated the isobaric core directly under the impactor by keeping the pressure constant for $h \le r_{imp}$ \citep{pierazzo2000understanding} . The pressure decrease according to $p \varpropto h^{-2}$ then starts in a depth of one impactor radius $r_{imp}$. We formally calculated the resulting volume filling factor over the whole depth of 100 km and then took the maximum value from before and after the considered impact. The size reduction of the target due to the compaction is not taken into account at this point, but is  applied after the last impact on each surface element. 4: Final random number, drawn from a quadratic distribution, determines where in the crater the considered element is located, with a value of zero meaning in the crater center and a value of unity meaning at the crater rim. 5: The observed surface element is cut out of the two-dimensional  cross section and reduced to a one-dimensional array from the surface to a depth of 100 km in 1-m steps. 6: This one-dimensional array is divided at the crater bottom into i) the crater ejecta, which either fall back and form a regolith layer or escape into interplanetary space forming meteoroids, and ii) the compacted remaining asteroid material. Information about the crater ejecta is saved, along with the experienced pressure and crater size, for later use. 7: The remaining surface element is filled up at the bottom with material that has the same volume filling factor as the lowest point so that the resulting array is again 100 km long and consists of $10^5$ elements. This then serves as the new surface element for further impacts, but with the pre-compaction of the earlier collisions. Information about the amount of material (i.e. the number of elements) ejected from the crater, i.e. the actual depth of the surface of the one-dimensional element, is stored for later use (see point 6 above). 8: In case of larger parent bodies, it is assumed that most of the material is recaptured by the parent body (see discussion in Sect. \ref{sec:intro}). This leads to the formation of a regolith layer on top of the compacted material. We assume that the ejected and re-accreted material is evenly distributed over the asteroid. Thus, the thickness of the layer accumulated per time step is calculated by the mean amount of total ejecta of the $10^4$ surface elements during 4.5 Gyrs, divided by the time span between two successive impacts. We assume that the re-deposited regolith particles form a layer with a volume filling factor of $\phi=0.6$ \citep{schraepler2015}. 9: As in step 2, the one-dimensional element with the record of the previous impacts is extended into a two-dimensional array, with a width corresponding to the crater radius of the subsequent impact.

\begin{figure*}[htp]
	\begin{center}		 \includegraphics[width=\textwidth]{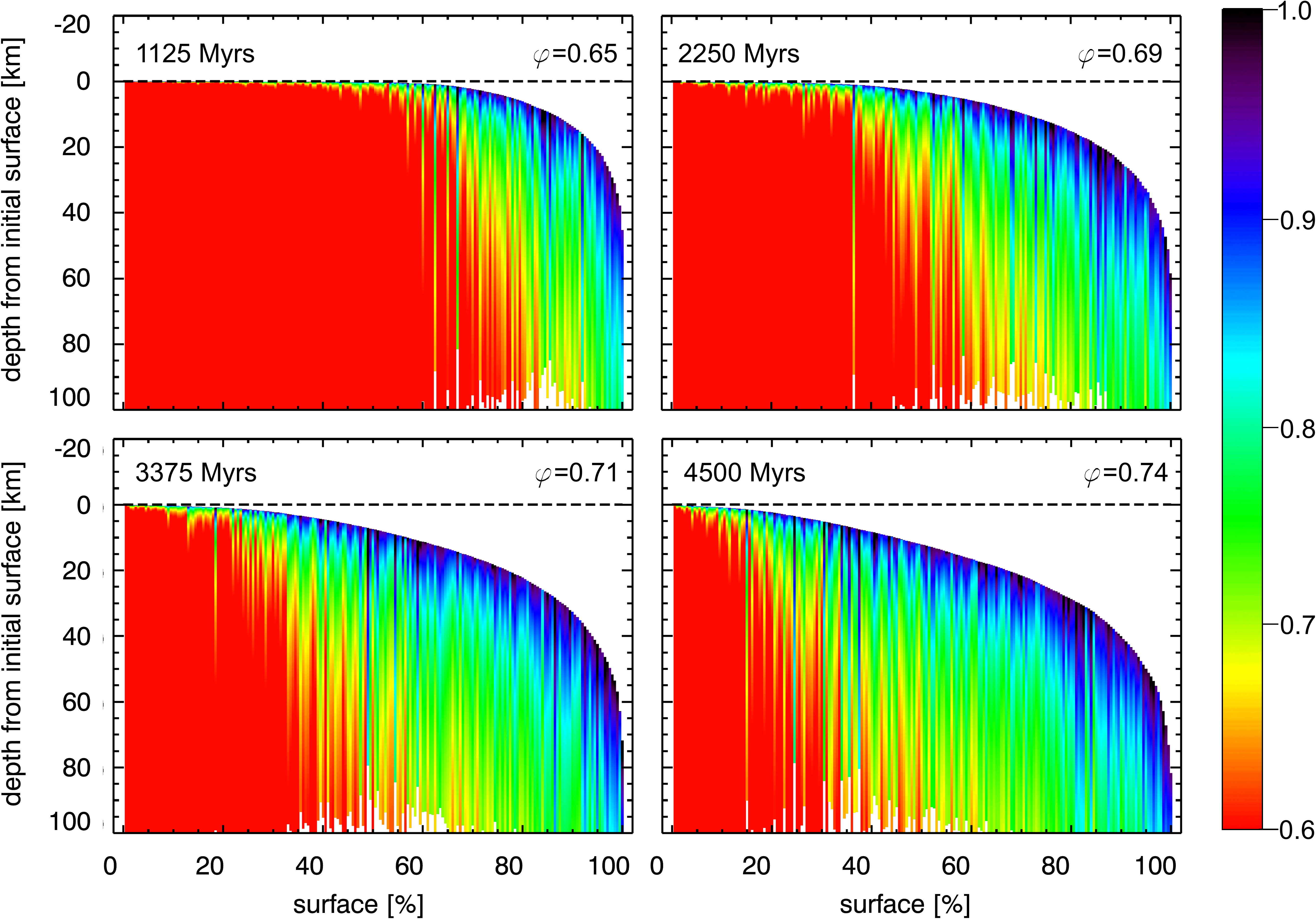}
		\caption{\label{fig:monte_movie}Snapshots of the temporal evolution of the cumulative crater depth and the interior compaction of an asteroid with initially 100 km radius for times as indicated in the upper left of the four panels. The colors denote the volume filling factors of 200 (arbitrarily chosen from the 10,000 available) one-dimensional elements, according to the color bar on the right, and sorted by the total crater depth. The plot show the compaction of the whole asteroid, and take the difference in depth from the initial surface i.e. the craters depth, and the compaction due to the higher volume filling factor into account. The occasional white lines at the bottom are caused by the material compaction. The horizontal dashed lines show the initial surface of the asteroid. The average volume filling factor of the compacted asteroid is indicated in the upper right of each panel. The initial volume filling factor of the whole asteroid was $\phi = 0.6$. Regolith re-deposition was neglected in this simulation.}
	\end{center}
\end{figure*}

To account for the smaller volume per unit mass for the impact-compacted material, the one-dimensional elements are compressed after the last impact. The size reduction factor is given by the ratio of the initial volume filling factor and the mean volume filling factor of the element. The distance to the initial asteroid surface (indicated by the horizontal dashed lines in Fig. \ref{fig:monte_movie} is calculated by moving the surface of the one-dimensional array downward by the sum of all crater depths experienced by this element. An example of a full simulation (without regolith deposition) is shown in Fig. \ref{fig:monte_movie}, where the four panels depict the collisional evolution of a simulated asteroid at four different times, as indicated in the upper left corner of each panel. The colors denote the volume filling factor (initially being $\phi=0.6$ throughout the asteroid), as shown by the color bar on the right. In this simulation, the collision rate is taken from the case of 5 $\rm km~s^{-1}$ impact speed, as shown in the bottom panel of Fig. \ref{fig:coll_number}. The surface elements are sorted by their cumulative crater depths. For visibility reasons, only 200 (out of 10,000) surface elements are shown. The average volume filling factor for each of the four times is shown in the upper right corner of the panels. This simulation is used as a reference to all other simulations.

\begin{figure}[htp]
	\begin{center}
		\includegraphics[width=8cm]{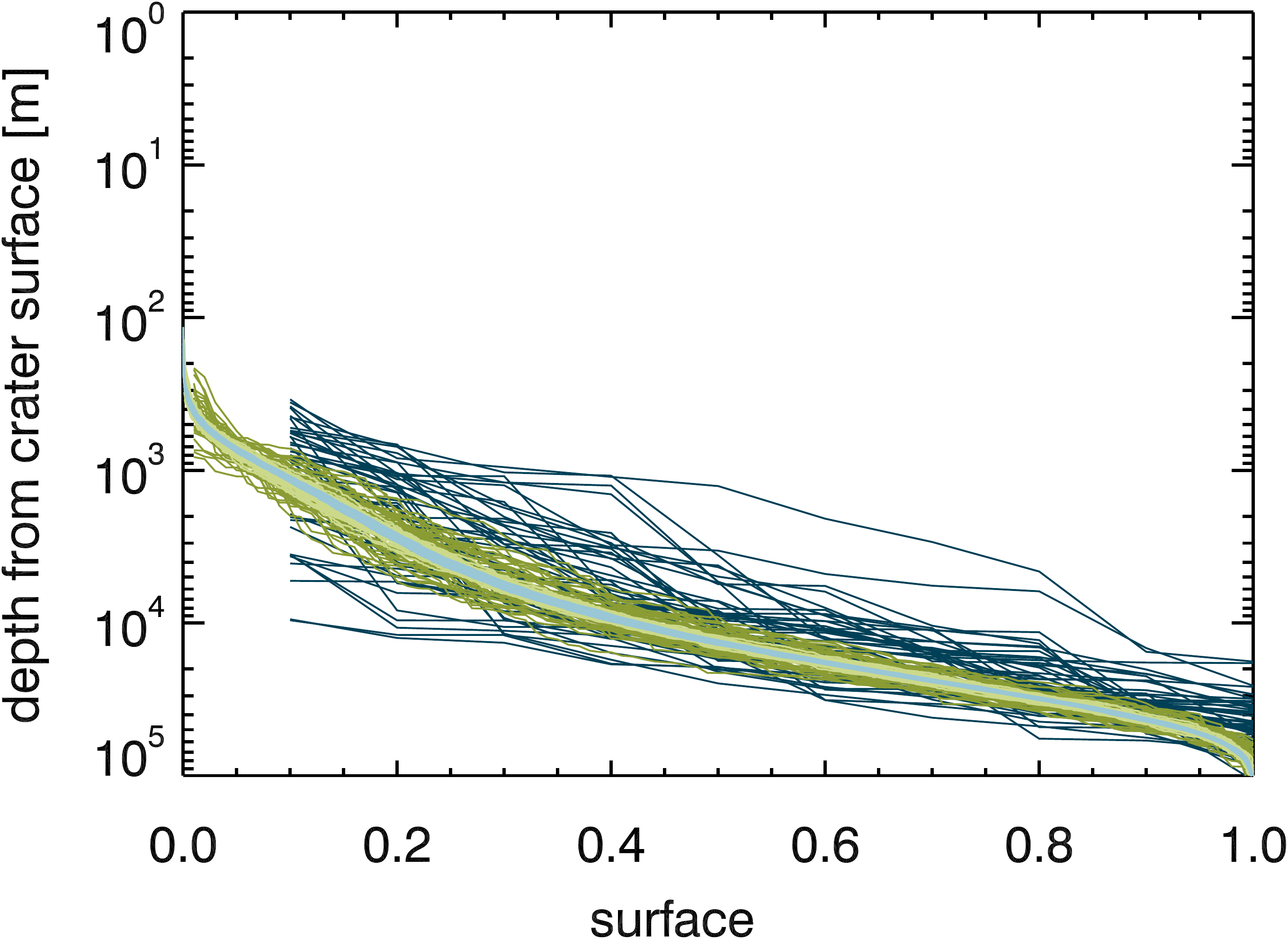}
		\caption{\label{fig:statistic_distribution} Spread of crater-depth distributions for 50 simulation with $10^1$ (dark blue), $10^2$ (dark green), $10^3$ (light green), and $10^4$ (light blue) one-dimensional elements, respectively. Mind that the ordinate is shown in logarithmic units.}
	\end{center}
\end{figure}

To gain statistical significance, in total up to $10^4$ one-dimensional elements were used in our Monte Carlo model. The total crater-depth distribution is shown in Fig. \ref{fig:statistic_distribution} for 50 different simulations with $10^1$ (dark blue), $10^2$ (dark green), $10^3$ (light green), and $10^4$ (light blue) surface elements, respectively. One can see that the spread between simulations with the same number of elements significantly decreases with increasing number of elements. For $10^4$ elements, the total spread in surface height is approximately $\pm 10$ percent, which we regard as being sufficiently precise.Contrary to intuition, the statistical significance does not significantly deteriorate towards the largest impactors, because our Monte Carlo method is based upon {\em crater areas} and not impact rates. As a comparison between the two panels of Fig. \ref{fig:coll_number} shows, the coverage for the largest impactor sizes is still several ten percent per size bin, whereas the total number of impacts is only on the order of unity for the full simulation time of 4.5 Gyrs.

Thus, although formally a single crater can only contain a single one-dimensional surface element (and large craters are rare), this is not a statistical problem in our code, because we calculated the probability distribution according to surface-area coverage so that effectively large craters are represented by accordingly more surface elements.

\section{Numerical simulations of asteroid compaction -- three-dimensional versus one-dimensional approach}\label{sec:1d_vs_3d}

The idea of this study is to predict the collisional evolution of an asteroid of 100 km radius over the lifetime of the Solar System through the simulation of the full sequence of experienced impacts and to compare the ejecta of recent collisions with typical chondritic meteorites. In this Section, we will show that a full three-dimensional model of asteroid compaction leads to the same results as a simplified one-dimensional model, which does not explicitly treat the spatial position of the impact. We will discuss the advantages and drawbacks of both models.

The first and most intuitive approach is to explicitly simulate all collisions that the asteroid experiences during its lifetime of 4.5 Gyrs. To realize this, we created a spherical three-dimensional asteroid and exposed it to a statistical bombardment randomly distributed over its entire surface. The parent body was discretized with a number of equidistant Cartesian surface points. The distance between two neighboring surface points determines the spatial resolution of the simulation. By choosing $10^5$ surface points, the minimum distance between two neighboring points is 1,016 m. Thus, a crater produced by an impact must be larger than this size to be detected by at least one of the surface elements. This minimum crater size corresponds to an impact of a projectile with at least $\sim 100$ m radius at an impact velocity of 5 $\rm km\,s^{-1}$. We chose as the minimum impactor radius 136 m, which produces a crater of 1,351 m radius. This ensures that more than one surface element is affected by each impact and no impact is lost between the surface elements. To resolve smaller impacts, the number of surface points has to increase quadratically with decreasing impactor radius, which requires immense storage and computing capabilities. For example, to resolve craters from impactors with a radius of only 0.1 m, the number of surface points has to be larger than $10^{11}$, with a total number of $10^{14}$ impacts to be explicitly computed, which is beyond our accessible computing power.

However, neglecting the very numerous small (and in terms of cratering relatively inefficient) impactors and choosing the range of projectile radii to fall between 136 m and 21,976 m, an explicit three-dimensional simulation of the impact history of a 100 km asteroid over 4.5 Gyrs of Solar System lifetime becomes doable. With these parameters, a statistical mean number of impacts of only 5,977 (see top panel of Fig. \ref{fig:coll_number}) has to be explicitly treated. We divided the total simulation time of 4.5 Gyrs into $10^6$ time steps of 4,500 years each to ensure that the probability of dual impacts per time step becomes negligible. For each time step, a list of random numbers is generated, to decide whether or not an impact of a particular impactor size from the chosen size range occurs during the considered time step. If an impact occurs, a random position of the surface is chosen to select where the impactor hits the asteroid. This results in a time sequence of impacts, represented by a list of impactors of different sizes randomly hitting the surface of the asteroid at a given position.

In our benchmark test of the three-dimensional simulations, the crater size only depends on the impactor size, because impact velocity and volume filling factor are kept constant for all impacts. Then, each impact is calculated sequentially according to the above-mentioned list. First, the linear distance of each (infinitesimally small) surface element from the point of impact is calculated. Surface elements that are closer to the point of impact than the calculated crater radius are selected for further treatment. For simplicity and for comparison with our one-dimensional model, we here (and only here) assume hemispherical crater shapes. For simplicity and only to compare the three-dimensional simulations to the one-dimensional mentioned below, we ignored compaction of the material below the crater floor.

\begin{figure}[htp]
	\begin{center}
		\includegraphics[width=8cm]{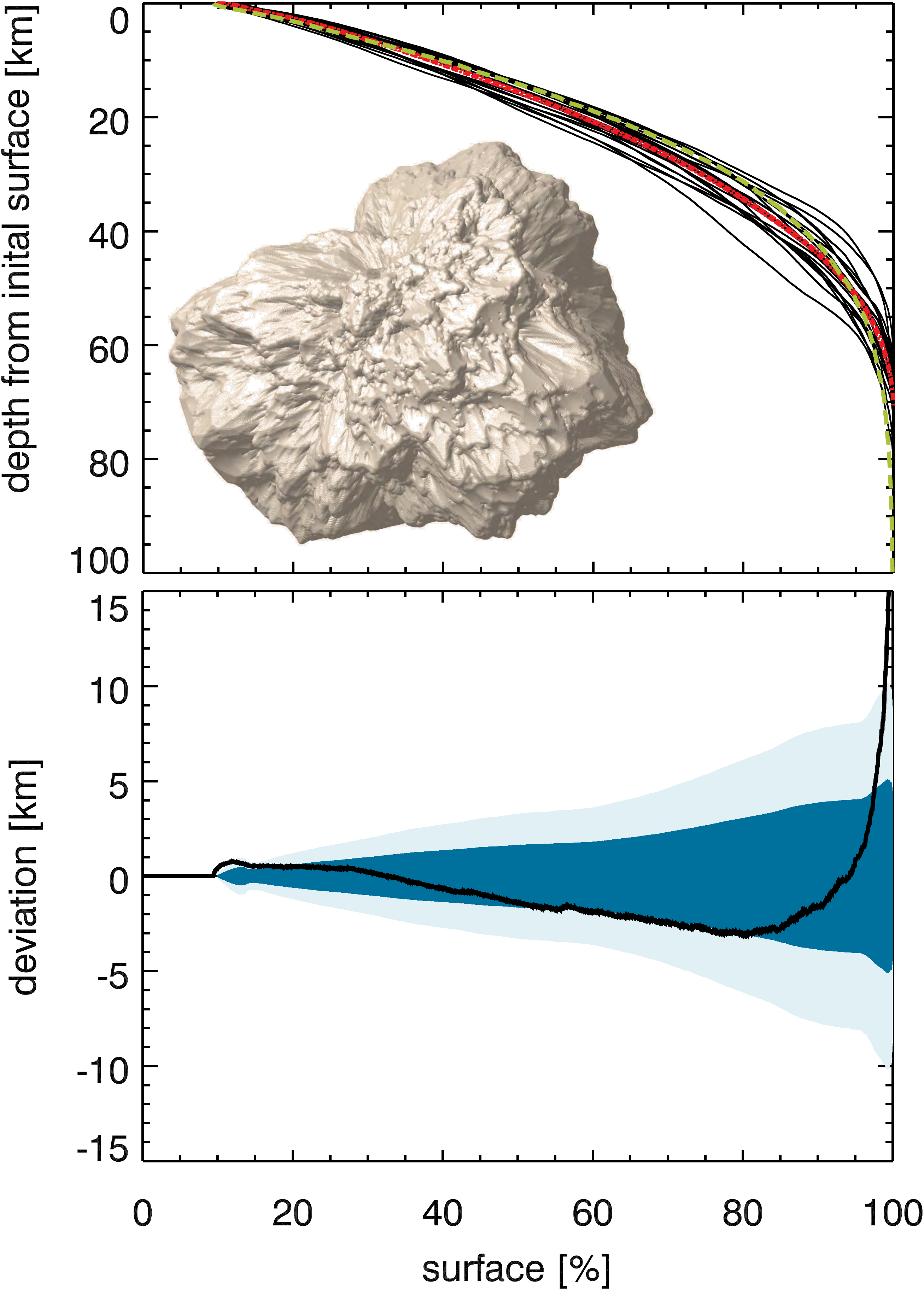}
		\caption{\label{fig:3d_vs_1d} Top: Comparison of the sorted crater depth distribution of 20 three-dimensional simulations using $10^5$ surface elements (shown as black solid curves) and the one-dimensional simulation with $10^4$ surface elements (shown as red dashed curve, see Fig. \ref{fig:statistic_distribution}). In both simulations, the size distribution of impactors was between 136 m and 22 km, the crater shape was assumed to be hemispherical and re-accretion of ejected material was neglected. The image in the inset shows the explicit result of the three-dimensional shape of one of the three-dimensional simulations with realistic crater shapes. A movie of the temporal evolution of an initially spherical body with 100 km radius due to cratering over 4.5 Gyrs can be found in the online version. Bottom: Statistical variations among the 20 three-dimensional simulations. The dark and light blue regions denote a range of one and two standard deviations from the mean, respectively. The black curve shows the deviation of the average one-dimensional from the mean three-dimensional profile.}
	\end{center}
\end{figure}

To get a statistically meaningful result, we ran the three-dimensional simulations 20 times, varying all random numbers. In the top panel of Fig. \ref{fig:3d_vs_1d}, we show the resulting 20 size-sorted distributions of the crater depths after 4.5 Gyrs of bombardment as black lines, along with a three-dimensional rendering of the asteroid shape. Here, one can see that about 10 percent of the surface have not experienced a considerable cratering, consistent in all 20 simulations. Due to small-number statistics, the larger craters show the widest spread among the different simulations, but follow the same trend. The impactors producing craters larger than the asteroid's size would lead to an overestimation of the crater depth in the outer regions of these craters in the 1-dimensional approach. This effect can be seen in Fig. \ref{fig:3d_vs_1d} as the 1-dimensional simulation exhibits larger craters on the right-hand side of the plot. However, as the bottom panel of Fig. \ref{fig:3d_vs_1d} shows, the mean one-dimensional depth profile falls within one standard deviation of the three-dimensional runs for about 95\% of the surface so that we consider the agreement between the two approaches as satisfactory.

If the explicit shape of the resulting object and the exact positions of the craters on the surface are not of interest, one can run the simulation in a one-dimensional fashion, as will be outlined in more detail in Sect. \ref{sec:Monte_Carlo_Code}. The one-dimensional method possesses several advantages over the explicit three-dimensional treatment. However, the one-dimensional simulations utilize an implicit method and can therefore only be interpreted in a statistical manner, because the individual surface elements do not correlate with each other and do not possess fixed positions on the asteroid surface.

In the one-dimensional case, the probability for a surface element to be covered by a crater is given by the mean crater coverage of the parent body, resulting from impactors with the size-frequency distribution of the asteroid belt as shown in the bottom panel of Fig. \ref{fig:coll_number}. Each surface element is covered with craters in a statistical manner with its position inside the crater also being randomly chosen for every impact. Thus, every surface element resembles the collision history for one asteroid.

Numerically, the advantage of the one-dimensional over the three-dimensional simulations is obvious, because no impact is lost between surface elements and the lower size limit of the impactors or crater sizes can be chosen arbitrarily small. Using the same size range of impactors as for the three-dimensional simulations, only two impacts occur on average on every surface element and 100 time steps suffice to ensure that a surface element is not hit more than once during the same time step. To gain a statistically meaningful average over many asteroids, $10^4$ surface elements were used. The effect of the number of surface elements is discussed in Sec. \ref{sec:simulation_parameters} and shown in Fig. \ref{fig:statistic_distribution}. The red dashed curve in Fig. \ref{fig:3d_vs_1d} shows the comparison between the one-dimensional and three-dimensional simulations. One can see that both methods agree very well with about 10 percent of the surface not being hit by any impact, in the overall slope of the crater-depth curve, and in the steep decline of the crater-depth curve at the upper end of the distribution. On top of this, the one-dimensional method allows to resolve craters from impactors with a radius of only 10 cm, as only the number of time steps must be small enough to prevent dual impacts within a single time step. Details follow in the next Section.

However, if computational power allows, the advantages of the three-dimensional model are obvious, namely a full representation of the simulated asteroid in all dimensions and the possibility to ``resolve'' craters that are larger than the mean spacing of the surface elements.

\section{Simulation results}\label{sec:results}

In this study, the Monte Carlo code described in Sect. \ref{sec:Monte_Carlo_Code} was applied to an asteroid with an initial radius of 100 km. An asteroid of this size is exposed to a bombardment of $\sim 10^{14}$ objects with radii between 0.1 m and 22 km during the lifetime of the Solar System (see Fig. \ref{fig:coll_number}). The total area covered by impact craters exceeds the asteroid's total surface area by a factor of $\sim 85-187$, depending on the mean impact velocity and the assumed  (see Fig. \ref{fig:coll_number}). This high surface coverage allows us to study the asteroid's temporal evolution in a statistical manner by assigning 10,000 representative surface elements and calculating the impact-driven evolution for each of them individually (see Fig. \ref{fig:statistic_distribution}). We ran two different kinds of simulations in which we either assumed that the material of the crater volume is being totally removed from the asteroid or fully re-accreted as loose regolith following the impact.

In this section, the results of the Monte Carlo simulations are analyzed for the effects of impacts on the physical appearance of the parent asteroid and with respect to the properties of the formed meteoroids. We will describe the differences between the performed simulations in terms of the total number of collisions, impactor size range and the effect of re-accretion of the ejected crater material.

\subsection{Simulation parameters}\label{sec:simulation_parameters}

In total we ran six individual simulations with an initial volume filling factor of 0.6, which are listed in Table \ref*{Tab:simulations}. The simulation mentioned in the first line is the reference simulation, covering the full range of mean impactor sizes of 0.13264 m and 21,976 m. For simplicity, these are referred to as 0.1 m and 22 km, respectively, in the rest of the paper. For the reference simulation, the collision number was chosen according to the higher mean velocity of 5 $\rm km\, s^{-1}$ (see Fig. \ref{fig:coll_number}). The same collision list for all 10,000 surface elements of this simulation, including impactor size, collision time and position within the crater, is used for all simulations with $N_{\rm col,5}$. By selecting only those surface elements from the simulations that never experienced an impact above a certain size limit, the effect of the largest impactors can also be studied using this simulation. Even for a maximum impactor radius of 1 km (instead of 22 km in the reference case), $\sim 2,500$ surface elements remain, which is statistically sufficient as shown in Fig. \ref{fig:statistic_distribution}. An additional advantage of using the same collision history in almost all simulations is that the influence of a specific parameter of the simulation on its results does not depend on the set of random numbers. Only for the runs with $N_{\rm col,3}$, a new set of random numbers was computed. The simulation F-HH-N was additionally used to study the influence of the initial volume filling factor on the collision outcome by varying it between 0.3 and 0.7. For the study of the influence of the small impactors (i.e. less than 1 m in size) on the volume filling factor and shape of the remaining asteroid, the simulation had to be run separately, because there is no surface element that does not experience such impacts.

\begin{table}[center]
\caption{Overview of the different simulations performed in this study. All simulations use the same collision list as described in the text. The initial volume filling factor in all cases was 0.6, except for F-HH-N, where we varied the volume filling factor between 0.3 and 0.7. The stated impact radii are the exact values used for the simulations, in the text they are referenced as LH (0.1 m - 22 km), HH (1 m - 22 km) and LL (0.1 m - 3 km).    }\label{Tab:simulations}
\begin{tabular}{c c c}
\small	
Case & Impactor radii $[\rm m]$  & Regolith \\
\hline
F-LH-N & 0.13264 - 21,976 & No \\
F-HH-N & 1.07 - 21,976 & No \\
F-LH-Y & 0.13264 - 21,976 & Yes \\
F-HH-Y & 1.07 - 21,976 & Yes \\
F-LL-Y & 0.13264 - 2,745 & Yes \\
\hline
\end{tabular}
\end{table}

\begin{figure}[htp]
	\begin{center}
	\includegraphics[width=8cm]{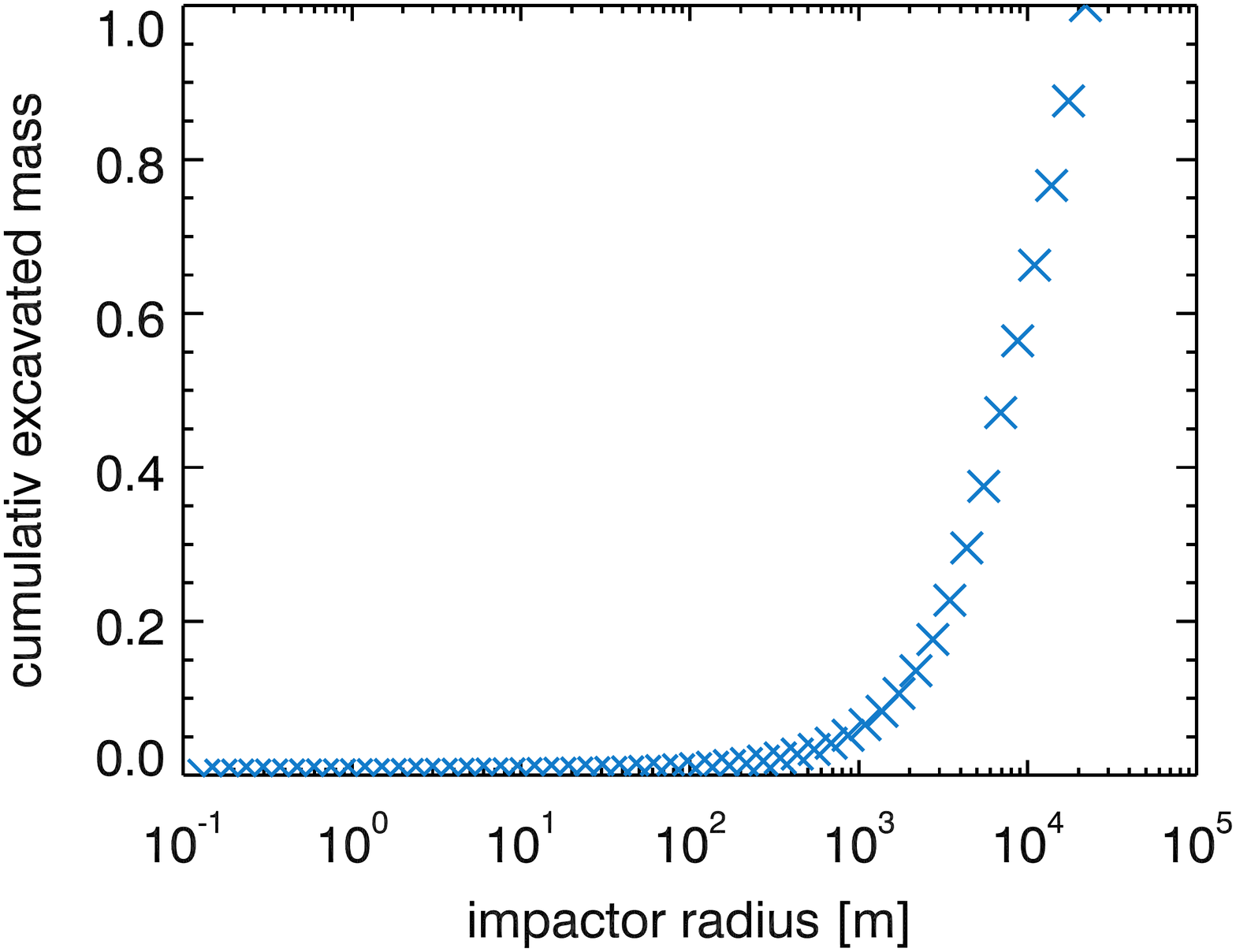}
	\caption{\label{fig:cum_exavated_mass} Normalized cumulative excavated mass as a function of mean impactor radius. Impacts follow the $N_{\rm col,5}$ distribution with the range of impact radii being 0.13264 m to 21,976 m.}
	\end{center}
\end{figure}

\begin{figure}[htp]

	\begin{center}
	 \includegraphics[width=8cm]{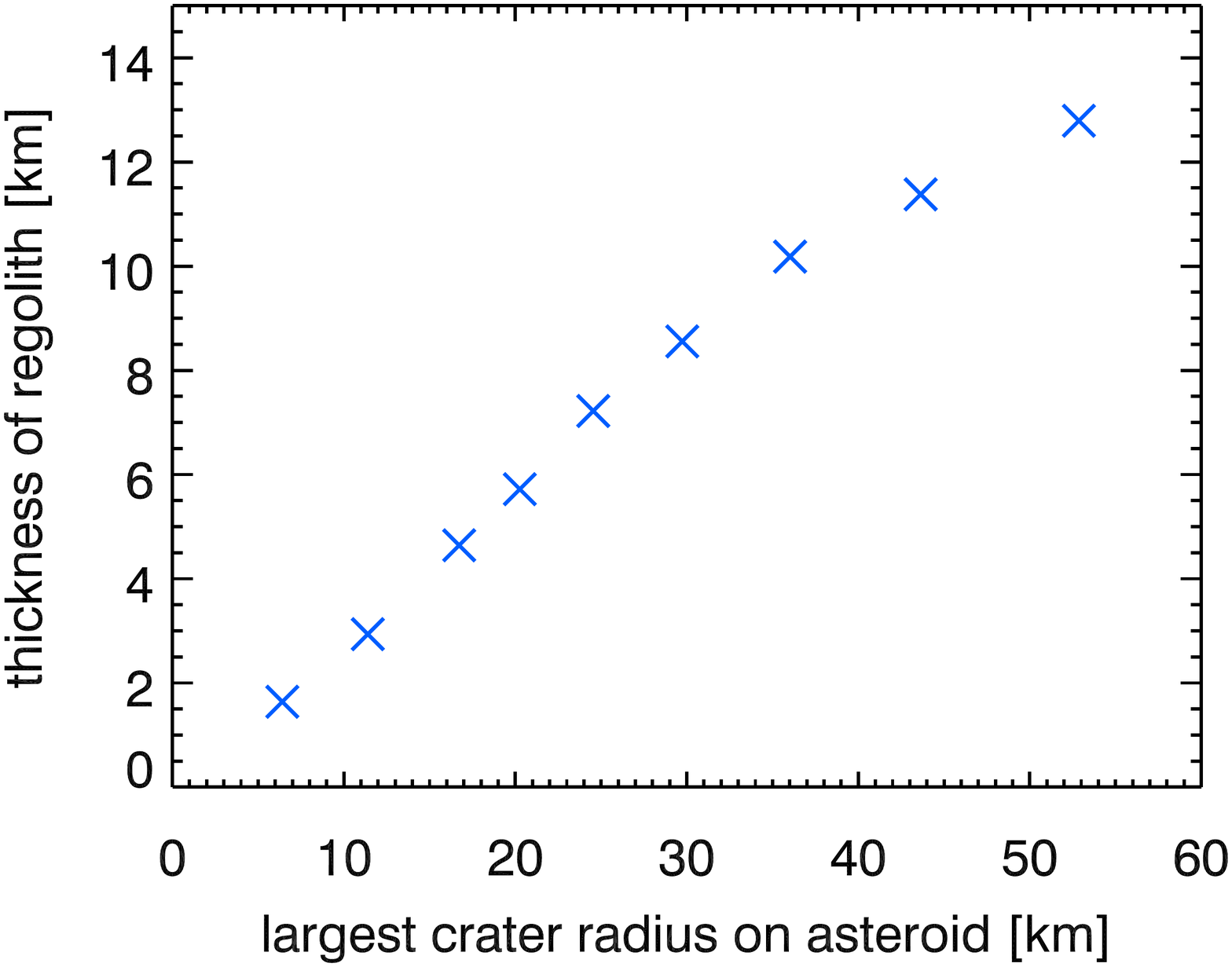}
	\caption{\label{fig:regolith_thickness_crater_size} Thickness of the regolith layer for an asteroid with a radius of 100 km as a function of the largest crater radius on the surface. The vertical line denotes the assumed largest possible crater radius of 100 km, which leads to a maximum regolith thickness of 18 km.}
	\end{center}
\end{figure}

The excavated mass fraction as a function of impactor size is shown in Fig. \ref{fig:cum_exavated_mass}. It is obvious that impactors above 1 km in radius are of utmost importance, as they account for more than 90 percent of the ejected mass. The mean height of the regolith layer according to all impactor sizes of the $N_{\rm col,5}$ probability distribution function was calculated to be 18 km for a 100 km sized asteroid (Fig. \ref{fig:regolith_thickness_crater_size}). Out of these, a thickness of only $\sim$ 80 m is formed by impactors with less than 1 m radius. Impactors of less then 3 km in radius account for 6.4 km of regolith deposit. Hence, the largest impactors (i.e. larger than 3 km in radius) are the most violent and account for a regolith layer with a thickness of about 11.6 km. It should be noted that the regolith thickness thus depends mostly on the infrequent impacts of the largest projectiles, which is taken into account in the simulation.

The effect of the largest impactors on the asteroid can be easily studied in the case without regolith by simply choosing surface elements that do not experience an impact with a projectile above a certain size limit. However, in the case with re-accretion of regolith, the total amount of excavated mass of all surface elements need to be known prior to the simulation. Thus, we ran two simulation with regolith deposition, one with the full distribution of impactor sizes (0.1 m - 22 km, F-LH-Y) and one considering only impactors of less then 3 km radius into account (F-LL-Y).

The effect of the impact velocity and the size of the impactors on the properties of the formed meteoroids, can be studied simply by choosing only the ejecta of those surface elements that were hit by impacts of a certain velocity or size range.

\subsection{Temporal evolution of the asteroid}
\begin{figure}[htp]
	\begin{center}
	\includegraphics[width=8cm]{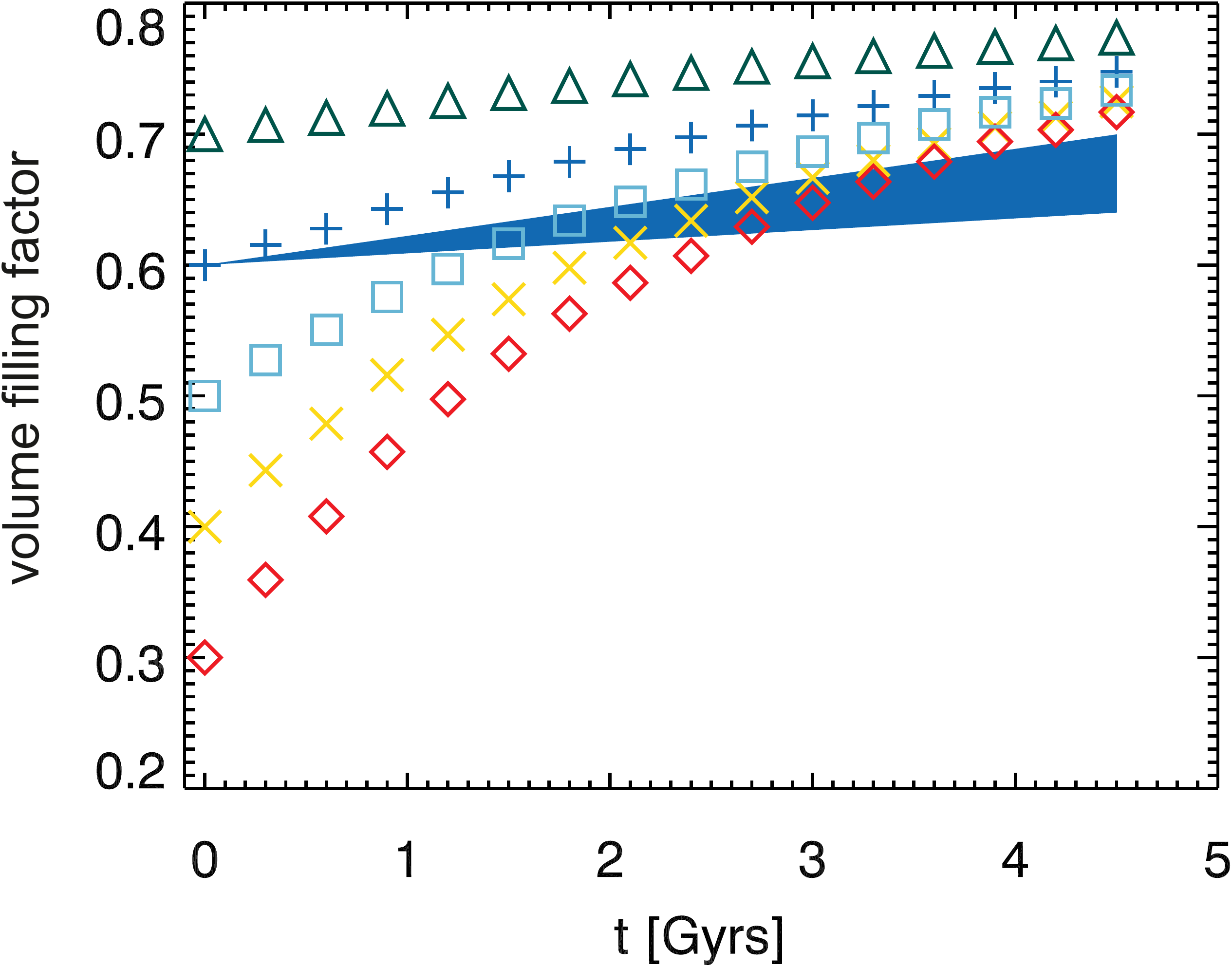}
	\caption{\label{fig:asteroid_compaction} Temporal evolution of the mean volume filling factor of asteroids with initial volume filling factors ranging from 0.3 to 0.7. All asteroids were exposed to the same bombardment with projectiles ranging in radius between 1 m and 22 km over 4.5 Gyrs (F-HH-N). The purple-shaded area shows the temporal evolution of an asteroid with initial volume filling factor of 0.6 with re-accretion of regolith. Here, the upper edge refers to impactors with radii between 0.1 m and 22 km  (F-LH-Y), whereas the lower edge is due to a bombardment of impactors with radii between 0.1 m and 3 km (F-LL-Y). }
	\end{center}
\end{figure}

Asteroids are believed to be among the most pristine objects of our Solar System. In this study, we try to assess how primitive in terms of collisional evolution asteroids really are. For this, we assume them to have once formed as spherical objects with an initial volume filling factor as a free parameter. Fig. \ref{fig:asteroid_compaction} shows the temporal evolution of the mean volume filling factor with initial volume filling factors ranging between 0.3 and 0.7 in the simulations F-HH-N. All simulated asteroids were exposed to the same constant bombardment of impactors with radii between 1 m and 22 km over a time period of 4.5 Gyrs. Due to the impact compaction, the difference in volume filling factor decreases over time and results in a final average volume filling factor of 0.87. The purple-shaded area in Fig. \ref{fig:asteroid_compaction} denotes the evolution in volume filling factor for simulations with re-accretion of regolith. Here, the upper edge of the blue-shaded area refers to impactors with radii between 0.1 m and 22 km (simulation F-LH-Y), resulting in a final volume filling factor of 0.82, whereas the lower edge is due to a bombardment of impactors with radii between 0.1 m and 3 km (F-LL-Y), which yields a final volume filling factor of 0.78.

Regolith re-accretion obviously reduces the average volume filling factor of the asteroid, because most impacts occur on the loose regolith and not on a pre-compacted surface. Thus, the impact pressure according to Eq. \ref{eq:pressure_melosh_easy} is lower, causing asteroids with regolith to exhibit a lower volume filling factor compared to those without. \citet{britt1987asteroid} estimated the mean volume filling factor of different asteroid types and found them to cluster around 0.8-0.9 (their Fig. 4). Thus their estimation is in general agreement with our simulations and falls exactly in between the volume filling factors of our simulations with and without regolith re-accretion. Information about the original (average) volume filling factor of the asteroid is deleted after 3 - 4 Gyrs in the case of no regolith. With regolith re-accretion, the volume filling factor increases linerly with time, with no saturation within the age of the Solar System. The final volume filling factor today is 0.8 and, thus, significantly lower than in the case without regolith and slightly depends on the size of the largest impactor (see Fig. \ref{fig:asteroid_compaction}). We conclude that any $\sim 100$ km asteroid with considerably lower volume filling factor than $\sim 0.8$ must be considered a rubble pile (which means that it has re-accreted after a catastrophic collision), whereas filling factors exceeding $\sim 0.9$ indicate internal melting. This is exactly what \cite{britt1987asteroid} concluded in their Figure 4.

\begin{figure*}[htp]
	\begin{center}		
    \includegraphics[width=\textwidth]{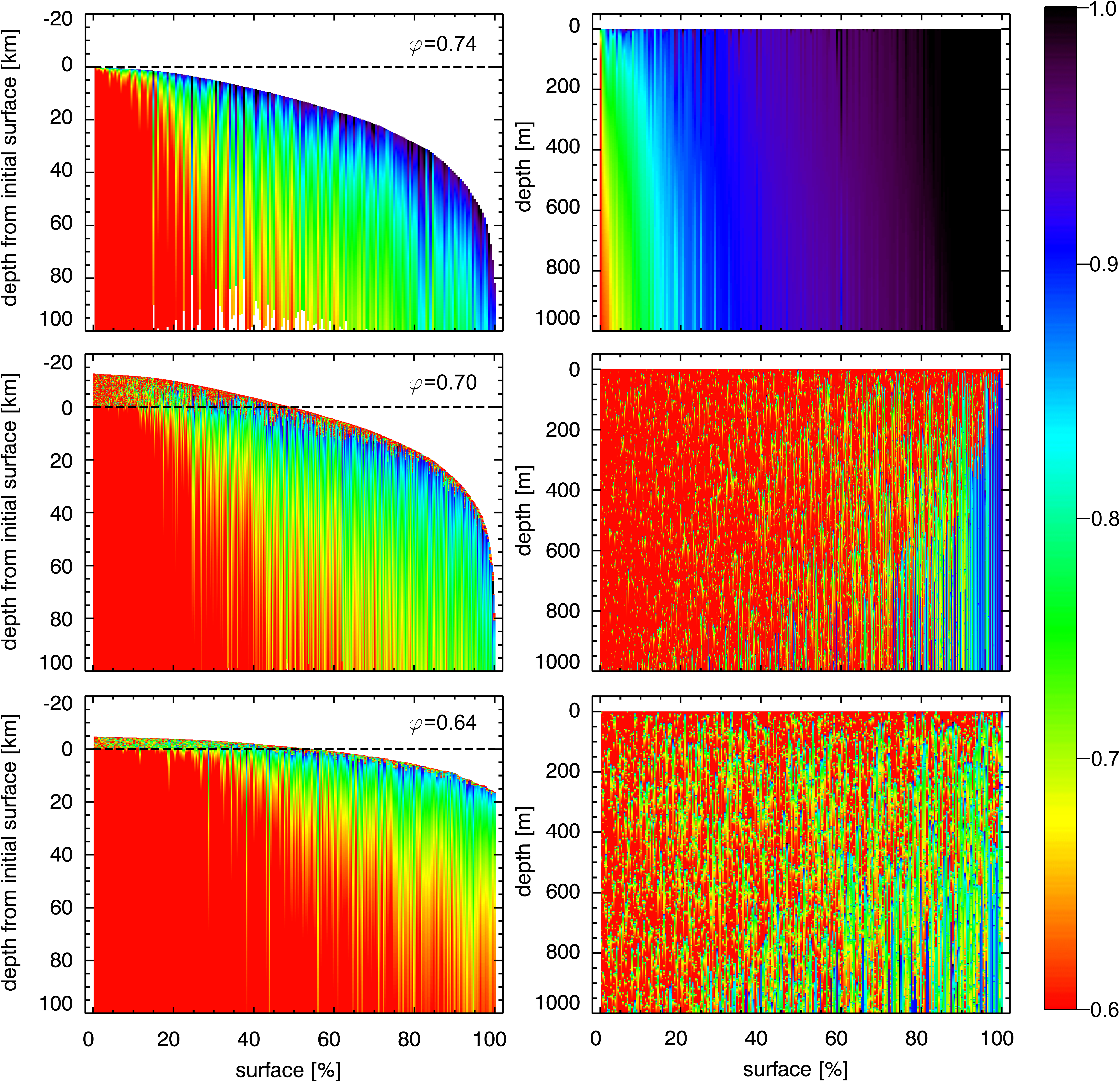}
	\caption{\label{fig:compaction_regolith}Asteroidal compaction after 4.5 Gyrs of continuous bombardment in identical notation as in Fig. \ref{fig:monte_movie}. The plots on the left-hand side show the compaction (color coded) and crater depth distribution (y-axis) of the whole asteroid, the plots on the right-hand side show a zoom into the uppermost 1,000 m of material. The top row shows the reference simulation (F-LH-N; the plot being identical to the one shown in Fig. \ref{fig:monte_movie} for 4.5 Gyrs), the center row refers to an asteroid with the same collision history, but with re-accretion of material (F-LH-Y). The bottom row indicates the simulation that ignores impacts with projectiles larger than 3,000 m (F-LL-Y). The horizontal dashed lines denote the position of the original surface of the asteroid. The value given in the upper right corner of the plots denote the average volume filling factor for the three cases. The initial volume filling factor in the simulation was 0.6.}
	\end{center}
\end{figure*}

In the following, we will focus on an initial volume filling factor of 0.6 and will present some more details on the internal structure of the asteroid after 4.5 Gyrs of collisional evolution. In Fig. \ref{fig:compaction_regolith}, we show, from top to bottom, the internal compaction for three cases, i.e. F-LH-N, F-LH-Y, and F-LL-Y, respectively (see Table \ref{Tab:simulations}). On the left-hand side of Fig. \ref{fig:compaction_regolith}, the full depth of the asteroid is shown, whereas on the right-hand side, we zoom in into the uppermost 1,000 m of the asteroid. The color bar represents the volume filling factor, ranging from values of 0.6 (red) to values of 1.0 (black). The sorting of the data with respect to the x-axis is according to the depth from the original surface.

Generally, the remaining asteroid is more compacted in its outer regions than close to the center, with the degree of compaction not being the same for all surface elements but varying significantly with the number and intensity of impacts encountered. The influence of regolith re-accretion on the overall compaction is very moderate but plays a major role for the upper layers of the asteroid (see below). Of more global importance is the presence or absence of large impactors, because they penetrate deep into the interior of the asteroid and cause significant compaction close to the center and deeper craters.

\begin{figure}[htp]
	\begin{center}		
    \includegraphics[width=8cm]{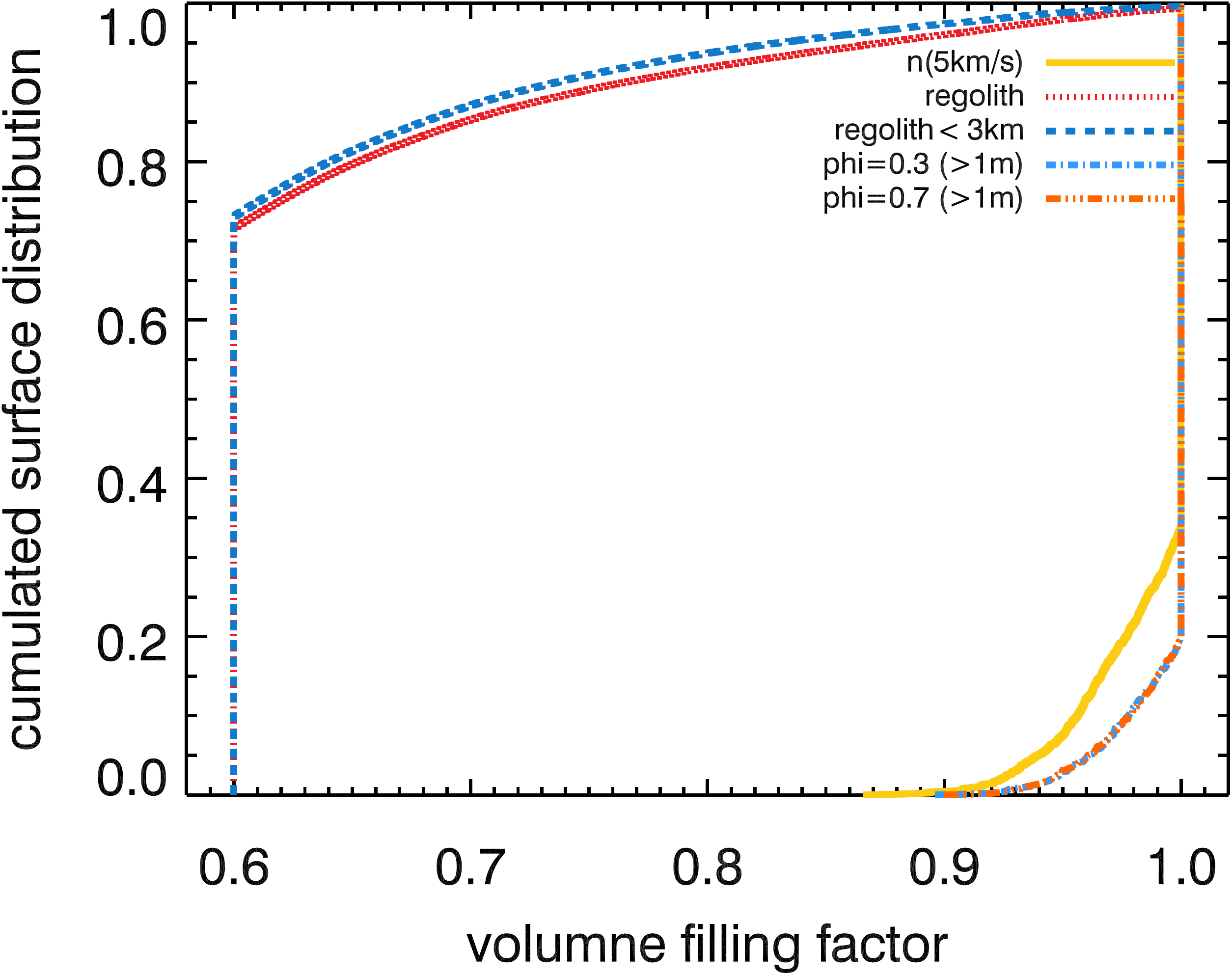}
	\caption{\label{fig:asteroid_surface_cum_oc_asteroid} Distribution of the average volume filling factor of the uppermost 10 m of the asteroid after a continuous bombardment of 4.5 Gyrs for an asteroid with and without re-accretion of regolith. Shown are the F-LH-N simulation (solid yellow curve), the F-HH-N simulation with initial volume filling factors of 0.3 and 0.7 (as dash-dotted light blue curve and dashed-dot-dot-dotted orange curve, both exactly lying on top of one another), respectively. The red dotted-dashed curve refers to those surface elements of the F-LH-Y simulation and the blue dashed curve to those of the F-LL-Y simulation that have only experienced impactors with radii of less than 3 km. The initial volume filling factor in these simulations was 0.6. }
	\end{center}
\end{figure}

The volume filling factor of the uppermost 1,000 m shows a clear difference if regolith re-accretion is considered or not. While the no-regolith case shows for most of the surface elements volume filling factors above 0.9, the regolith re-accretion forms a roughly 1 km deep layer with a volume filling factor 0.6, which is then only slightly compacted by subsequent impacts. However, a compacted crust is also formed in this case, which can be seen by considering the average volume filling factor of the uppermost 10 m of the asteroid as shown in Fig. \ref{fig:asteroid_surface_cum_oc_asteroid}. As the simulation without regolith (F-LH-N) exhibits a very compacted surface material with a mean volume filling factor of more than 0.9 and $\sim 70$ percent of the surface even possessing no porosity, the simulations with regolith (F-LH-Y and F-LL-Y) are dominated by the re-accreted material. About 70-75 percent of the surface element have not been compacted at all in the uppermost 10 m, with the rest having experienced compression by small impactors. This higher degree of compaction is consistent with the fact that a significant fraction ($\leq 15 \%$) of chondrites are regolith impact breccias that can be considered as being a by-product of this collisional processing \citep{bischoff2006nature}.

From Fig. \ref{fig:compaction_regolith} it can be seen that re-accretion of material leads to a regolith thickness of $\sim 18$ km for the full range of projectile sizes (simulation F-LH-Y), but only to a thickness of $6.4$ km for the case of a reduced upper projectile size (simulation F-LL-Y). The influence of the size of the largest projectile on the overall volume filling factor of the entire asteroid as well as for the uppermost 1,000 m is rather small. The average volume filling factor decreases from 0.82 to 0.78, when the largest projectile size is reduced from 22 km to 3 km. Without re-accretion of material, the average volume filling factor is 0.88. For the uppermost 1,000 m, the differences in the two cases with regolith re-accretion are also small, with the thickness of the uncompressed layer being roughly half as thick in the case of the largest projectile being 3 km in radius.

\subsection{Properties of the forming meteoroids}
\begin{figure}[htp]
	\begin{center}
		\includegraphics[width=8cm]{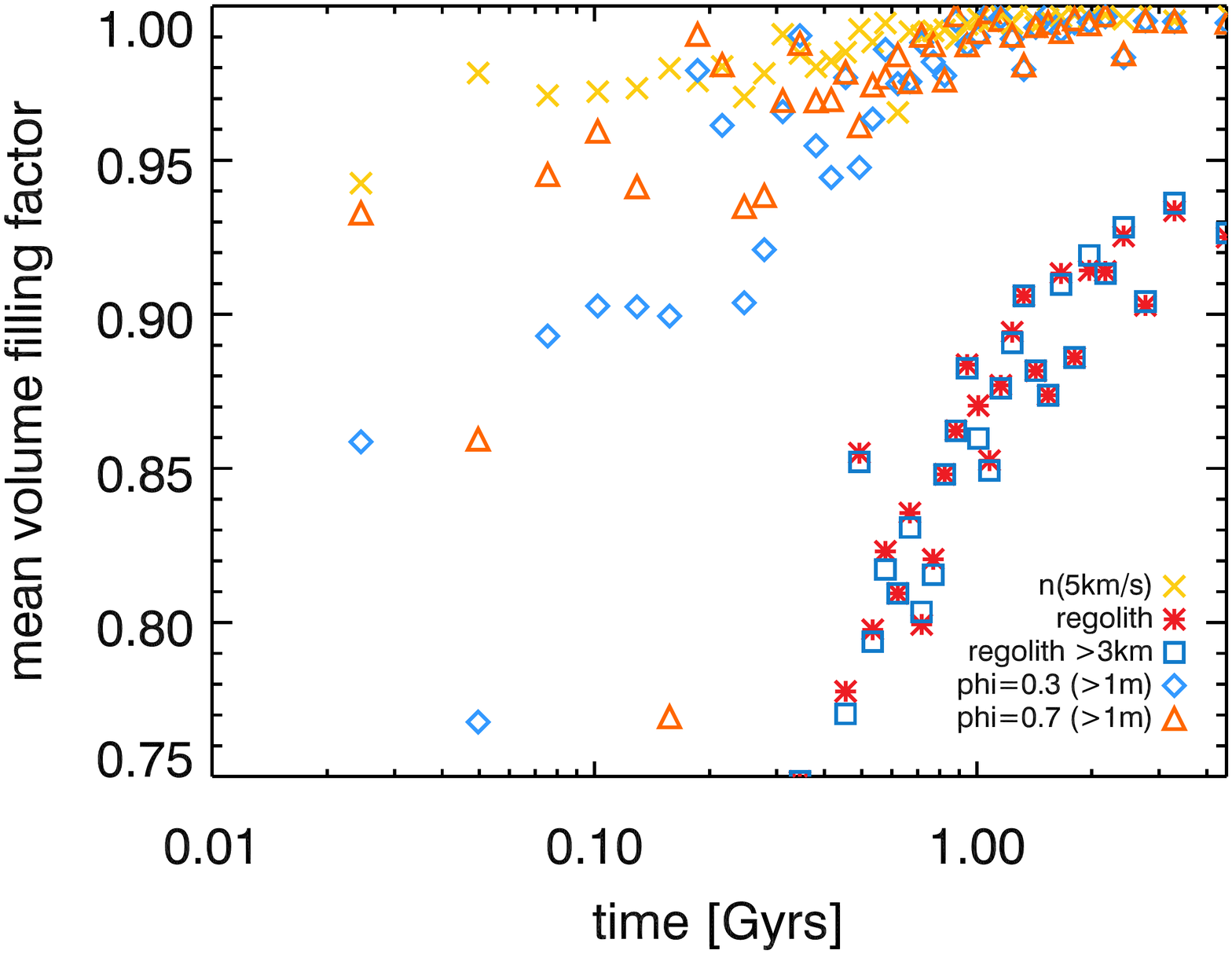}
		\caption{\label{fig:met_time_evoltuion} Mean volume filling factor of the ejected meteoroid mass as a function of time after the formation of the asteroid. Besides the reference case (F-LH-N, crosses), initial volume filling factors of 0.3 (diamonds) and 0.7 (triangles), respectively, are shown. In these cases, the smallest impactor size was set to a radius of 1 m (F-HH-N). Two regolith cases are denoted by asterisks (F-LH-Y) and squares (F-LL-Y), which differ in the maximum projectile size of 3 km and 22 km, respectively.}
	\end{center}
\end{figure}

As described in the introduction, meteorites are the best source of available asteroidal material. Most of them show an absolute age of 4.5 Gyrs, and a CREA of $\sim 20$ Myrs. To make the crater ejecta of our simulation comparable to the contemporary meteorites, a time span of 20 Myrs is used as temporal bin size. Due to the impact compaction of the underlying asteroid, the pressure in subsequent impacts is higher than in previous impacts and therefore also the volume filling factor of the crater ejecta increases over time. Fig. \ref{fig:met_time_evoltuion} shows the mean volume filling factor of the formed meteoroids in 20 Myrs time steps at different times after the formation of the asteroid. For comparison, we show the results of the reference simulation (F-LH-N, crosses) and the simulation F-HH-N with initial volume filling factors of 0.3 and 0.7, respectively (diamonds and triangles), all three without re-accretion of regolith. Furthermore, two simulations in which the ejected material is re-accreted as regolith, are shown by asterisks (F-LH-Y) and squares (F-LL-Y). The mean volume filling factor of the formed meteoroids in the reference simulation reaches a constant value of $\sim 0.98-1$ after only $\sim 500$ Myrs. This fast surface-compaction process is dominated by the smallest impactors, which can be seen when comparing the temporal evolution of the compaction with  simulations using different initial volume filling factors and allowing only impactors larger than 1 m. This decreases the total surface coverage from 187 to only 11 for the $N_{\rm col,5}$ case. Even if the simulation starts with an initial volume filling factor of 0.3, the meteoroids formed in the simulation obtain an almost compact state after only $\sim 500$ Myrs, which is the average time in which each surface element of the simulation is hit once and after which the difference between the initial volume filling factor of 0.3 and 0.7 has vanished. If re-accretion of regolith is taken into account, the impactors most likely hit a surface element covered by regolith, which leads to a steady temporal increase in volume filling factor, with a present value of $\sim 0.93$. A difference induced by the largest impactor size could not be found in the regolith case.

\section{Discussion}\label{sec:discussion}
In this section, we will interpret the results presented in the previous section. First, we will discuss the expected internal structure of the asteroid and the influence of the largest impactor on the internal structure and excavated mass. Then the height profile of our simulations is compared to that of the asteroid (21) Lutetia. Hereafter, we will discuss the required properties that lead to a maximal pressures experienced by the crater ejecta, which resemble the measured shock stages of the excavated meteorites. Finally, we will propose a model for the evolution of asteroids and the formation of meteoroids.

\subsection{Internal structure of primitive asteroids}
As shown in Fig. \ref{fig:compaction_regolith}, our simulations reveal the internal porosity of the contemporary asteroids and the influence of the re-accretion of regolith. Although the overall volume filling factor only decreases from 0.88 (case F-LH-N) to 0.82 (case F-LH-Y) when a regolith surface is considered, the internal structure is quite different. The most remarkable difference is unsurprisingly found within the $\sim 18$ km of accreted regolith. Due to the continuous impacts onto pre-shocked material, the case of no regolith accretion exhibits about 70 percent of the uppermost 1 km to be completely non-porous (see top right panel in Fig. \ref{fig:compaction_regolith}). On contrast, with re-accretion of regolith in between subsequent collisions, the porosity of the uppermost 1 km layer is determined by the deposition process and, thus, obtains values of $\sim 0.6$. The presence of a regolith layer is supported by multiple observations of the shock stages and annealing features experienced by ordinary chondrites. \citet{rubin2004postshock} envisioned that ordinary chondrite annealing was consistent with rocks buried beneath the floor or lining the walls of an impact crater or deposited in a hot, thick ejecta blanket.

Although the surface coverage by impact craters is much higher for small projectiles (see Fig. \ref{fig:coll_number}), the largest impactors determine the thickness of the regolith layer, as can be seen by comparing the middle and the bottom panel of Fig. \ref{fig:compaction_regolith}. Reducing the upper limit of the projectile size distribution from 22,000 m radius to 3,000 m (case F-LL-Y) leads to a further decrease in volume filling factor to 0.78 and a decrease of the regolith thickness to only $\sim 6.4$ km. The intensity of compaction that reaches the center of the asteroid systematically decreases from the case with no re-accretion of regolith to the case with re-accretion of regolith and a reduced upper projectile size. We conclude that the most primitive (un-compacted and un-shocked) ordinary chondrite like material can only be found in the deep interior of asteroids, which have a key implication for envisioned sample return missions. However, carbonaceous asteroids could behave differently as they contain volatiles and are more matrix rich.

Our simulations suggest that the moment of inertia of primordial asteroids should be slightly higher than that of homogeneous bodies, because the average compaction increases radially outward. In the case of a spherical asteroid, we expect the moment of inertia to be $I > (2/5) M R^2$, with $M$ and $R$ being the mass and radius of the asteroid. Due to uncertainties in the spatial distribution of the largest craters, however, we cannot be more quantitative at this point.

\subsection{Influence of the maximum impactor size}
As we have shown in Fig. \ref{fig:cum_exavated_mass}, the majority of the excavated mass stems from a few largest impactors. Here the question arises how probable the collision with an impactor within a given size range is. To calculate the probability $P_k(T)$ for a target body to be hit $k$ times over a time period $T$, we use the Poisson statistics of independent events, i.e.,
\begin{equation}\label{poisson1}
P_k(T,r_{\mathrm{min}},r_{\mathrm{max}}) = \frac{\lambda^k}{k!} \exp{(- w(r_{\mathrm{min}},r_{\mathrm{max}})T)} .
\end{equation}
Here, $w(r_{\mathrm{min}},r_{\mathrm{max}})$ is the collision rate per unit time of the target with impactors of the size between $r_{\mathrm{min}}$ and $r_{\mathrm{max}}$. The above expression simplifies to
\begin{equation}\label{poisson2}
P_0(T,r_{\mathrm{min}},r_{\mathrm{max}}) = \exp{(- w(r_{\mathrm{min}},r_{\mathrm{max}}) T)}
\end{equation}
for the case of no collision, i.e. $k=0$. In Table \ref{probability}, we summarize $P_0(T,r_{\mathrm{min}},r_{\mathrm{max}})$ for $T=4.5$ Gyrs. As can be seen in Table \ref{probability}, a target with 100 km radius is very unlikely to escape an impact of bodies with at least 6 km radius, whereas a target body half its size (50 km radius) possesses a $\sim 20$ percent chance to do so. Thus, we varied the maximum size of the impactors in order to investigate their quantitative influence on the collision outcome.

\begin{table}
	\centering
		\caption{Probability of a target body to escape impacts with projectiles in the size range between $r_{\mathrm{min}}$ and $r_{\mathrm{max}}$ over the age of the Solar System ($T=4.5$ Gyrs). Mind that the maximum impactor radius of 24.67 km is the upper end of the logarithmic bin centered at 21.98 km. }\label{probability}
	\begin{tabular}{rrr}
		Target radius (km) & $r_{\mathrm{min}}$-$r_{\mathrm{max}}$ (km) & $P_k(T,r_{\mathrm{min}},r_{\mathrm{max}})$\\
		\hline
		100 & 19.58-24.67 & 8.1E-1\\
		100 & 15.54-24.67 & 6.5E-1\\
		100 & 12.33-24.67 & 4.3E-1\\
		100 & 9.79-24.67 & 1.7E-1\\
		100 & 7.77-24.67 & 3.9E-2\\
        100 & 6.16-24.57 & 1.7E-3\\
        100 & 4.89-24.57 & 1.2E-5\\
        100 & 3.88-24.57 & 5.3E-10\\
		\hline
		50 & 19.58-24.67 & 9.5E-1\\
		50 & 15.54-24.67 & 9.0E-1\\
		50 & 12.33-24.67 & 8.1E-1\\
		50 & 9.79-24.67 & 6.4E-1\\
		50 & 7.77-24.67 & 4.5E-1\\
        50 & 6.16-24.57 & 2.0E-1\\
        50 & 4.89-24.57 & 5.9E-2\\
        50 & 3.88-24.57 & 4.8E-3\\
		\hline
		\end{tabular}

		\end{table}

From the previous sections, it has become clear that the maximum impactor size is of utmost importance for the collision outcome and the properties of the resulting meteoritic material. Fig. \ref{fig:regolith_thickness_crater_size} shows the resulting regolith thickness as a function of the radius of the largest crater on the surface, which is a measure of the size of the largest impactor. Although the data shown in Fig. \ref{fig:regolith_thickness_crater_size} refer to an asteroid with 100 km radius, this should be also valid for asteroids of different sizes as long as almost all excavated material is re-accreted as regolith, because the coverage of the surface is independent on the asteroid size (see Fig. \ref{fig:coll_number}). We find an almost linear increase of the regolith thickness with increasing crater size up to a layer depth of $\sim 15$ km, and a somewhat shallower increase for larger crater sizes. Thus, the measurement of the size of the largest crater might be used for the determination of the total depth of regolith on an asteroid.

\subsection{Surface profile of asteroids}

In this subsection, we will compare the surface profile of our model asteroid, as shown in Fig. \ref{fig:compaction_regolith}, with the surface profile of asteroid (21) Lutetia, because this asteroid has a size similar to the one used in our simulation and a high-resolution shape model is available \citep{Farnhamlutetiashape}. We assume Lutetia to be once formed as a spherical object around with its center of mass identical to the present. Furthermore, we assume that the largest distance of the surface to the center of mass of (21) Lutetia is the initial radius of the asteroid. Applying the shape model by \citet{Farnhamlutetiashape}, we then derived the surface profile of (21) Lutetia by calculating the distance of each surface element of the shape model to the center of mass. As Lutetia is a large body with a maximum diameter of 121 km, it re-accreted almost all material produced by impacts during its lifetime \citep{o1985impact}.

\begin{figure}[htp]
	\begin{center}
		\includegraphics[width=8cm]{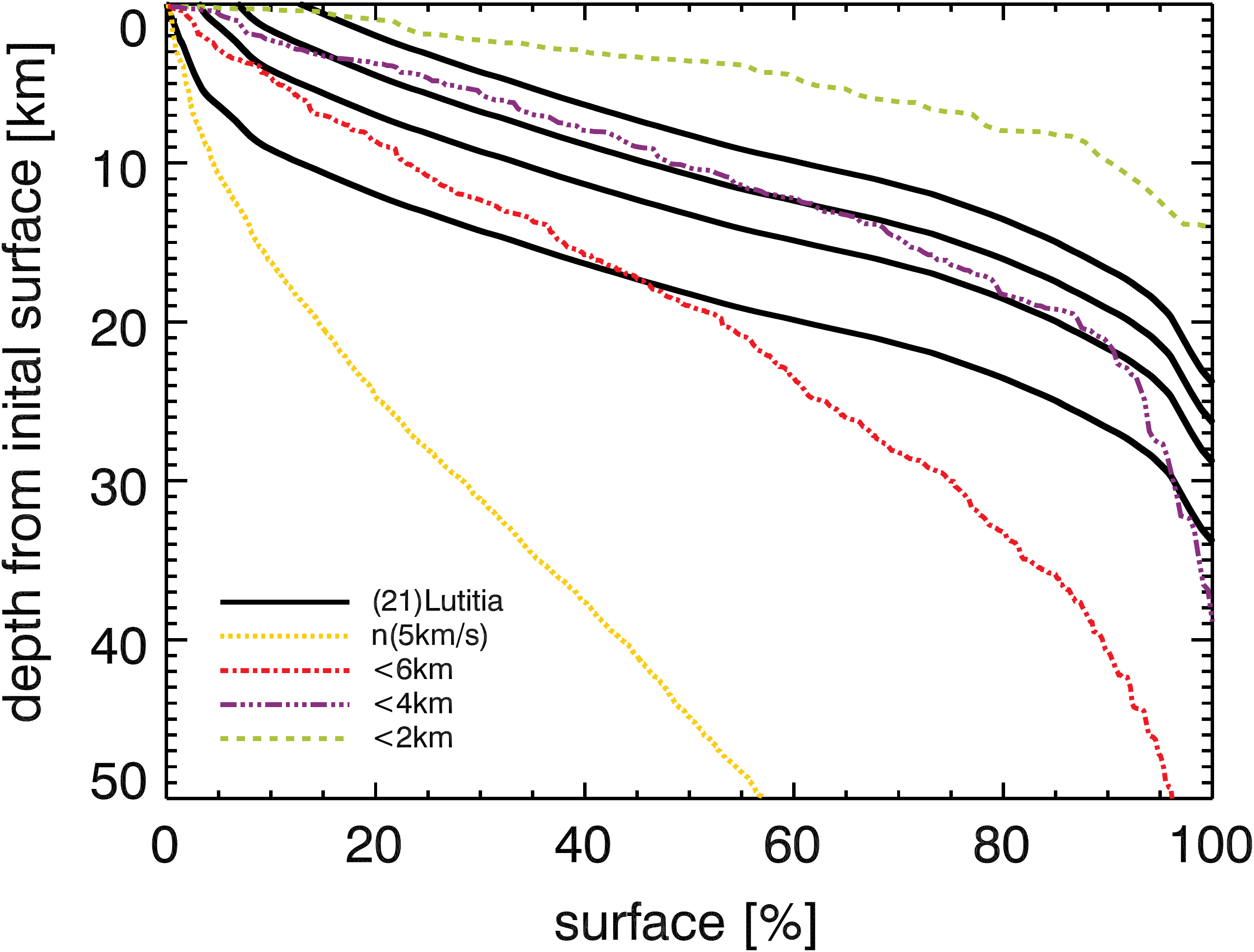}
		\caption{\label{fig:statistic} Comparison of the surface profiles of asteroid (21) Lutetia (based on the shape model of \citet{Farnhamlutetiashape}) and our simulations. The four solid curves for asteroid (21) Lutetia are, from bottom to top, the actual surface profile and the surface profiles shifted upwards by 5, 7.5 and 10 km, respectively. The results of four simulations are labeled by n(5km/s) for the full simulation F-LH-Y and by $<6$km, $<4$km, and $<2$km, in which the maximum impactor size was restricted to 6 km, 4 km, and 2 km, respectively.}
	\end{center}
\end{figure}

From Fig. \ref{fig:regolith_thickness_crater_size}, we can infer the thickness of the regolith layer on (21) Lutetia to be around 4 km, because the largest crater on (21) Lutetia has a diameter of about 55 km \citep{sierks2011images}. In Fig.  \ref{fig:statistic}, the actual surface profile of (21) Lutetia is shown as the lowermost of the four solid curves. It is obvious that Lutetia's overall surface profile cannot be reconstructed with our full model F-LH-Y (labeled n(5km/s) in Fig. \ref{fig:statistic}).

However, the much shallower decline of the depth profile of asteroid (21) Lutetia and the size of its biggest crater suggest that Lutetia was not hit by projectiles all the way up to more than 20 km in radius, as we used in simulation F-LH-Y. Reducing the maximum impactor radius to 6 km, 4 km, and 2 km, respectively, results in the three other simulation curves shown in Fig. \ref{fig:statistic}. Furthermore, from the profile of the highest 10 percent of Lutetia's surface, it is obvious that the asteroid possesses some kind of ``mountain'' with about 10 km in height (see below). Thus, we ignored the most elevated parts of (21) Lutetia by flattening them by 5, 7.5 and 10 km, respectively. The results are shown as the other three solid lines in Fig. \ref{fig:statistic}. A comparison of these surface profiles of asteroid (21) Lutetia with our simulations shows quite good agreement for a maximum impactor size of $\sim 4$ km in radius. We find both, the slope of the surface profile in the central $\sim 60$ percent of the surface as well as the presence and depth distribution of the deepest $\sim 20$ percent of the surface reasonably well represented by our model.

\citet{Vincent201279} measured the depth-to-diameter ratio for Lutetia's craters and found values ranging from 0.05 to 0.3, with the higher values being related to the youngest craters. Thus, the difference in the surface profile for the very largest craters (and, thus, the deepest terrain) can be explained by a difference in their formation time between the simulation and asteroid (21) Lutetia, because the relation between shape and age of the craters can be explained by the continuous re-accretion of regolith and intrinsically slightly shallower craters, due to back-flowing material from the crater walls towards the crater center. This may explain why the slope of the deepest $\sim 10$\% of the terrain is somewhat steeper in our simulations.

The above-mentioned ``mountain'' on asteroid (21) Lutetia is obviously a realistic feature. The furthest point on Lutetia's surface is at a distance of 67.770 km from its center of figure. Displaying the uppermost 10 km on Lutetia (corresponding to 12.8\% of Lutetia's surface) shows that they belong to only three individual geographical regions (i.e. three ``mountains´´). The uppermost 7.5 km (6.8\% of Lutetia's surface) belong to two ``mountains'' and the uppermost 5 km (3.1\% of Lutetia's surface) belong to just one ``mountain''.

\subsection{Comparison between the maximum experienced pressure and the measured shock stages of meteorites}
After a bombardment of 4.5 Gyrs duration, our two simulation types, i.e., with and without regolith, exhibit significantly different distributions of the maximum pressure and, thus, the shock stage of the produced meteoroids. Here, we will compare the simulated shock stages with those of real meteorites, which certainly must be done with caution. For this comparison, we used all ordinary chondrites from the Meteoritical Bulletin Database that possess an unambiguous assignment to one of the shock stages defined by \citet{StoefflerEtal1991}. As these are only defined within different distinct pressure ranges with some gaps in between, but the pressures derived in the simulation is continuous, the transition range between two neighboring shock stages will be dedicated to both, the lower and the higher shock stage. We only use ordinary chondrites with a petrologic type less than 4, because these are the least thermally altered meteorites, which most likely stem from undifferentiated asteroids or from the surface of differentiated parent bodies. If the parent body is differentiated, its volume filling factor must be higher in the internal region \citep{Henke2012b}, which leads to higher pressures and higher shock stages. In order to simulate differentiated parent bodies, the thermally compacted material must be tracked in the simulations, which was not done in this study and remains as future work. It must be clear that the data from the Meteoritical Bulletin Database do by far not contain all known meteorites, nor do all of the listed meteorites have an unambiguously assigned shock stage. Additionally, we do not distinguish between the three main groups of ordinary chondrites, which stem from different parent bodies. We assume that the material of the L, LL, and H chondrites behave not too different when impacted. As each of our simulations is an average of 10,000 single one-dimensional surface elements, we assume that an averaging over the three chondrite groups (L, LL, H) should be consistent with our approach. Thus, each surface element either represents an individual target body or an arbitrary point on the surface of that body. Each surface element has its own collision list and the entirety of all 10,000 surface elements represents an ensemble of bodies or a single body.

\begin{figure}[htp]
	\begin{center}
		\includegraphics[width=8cm]{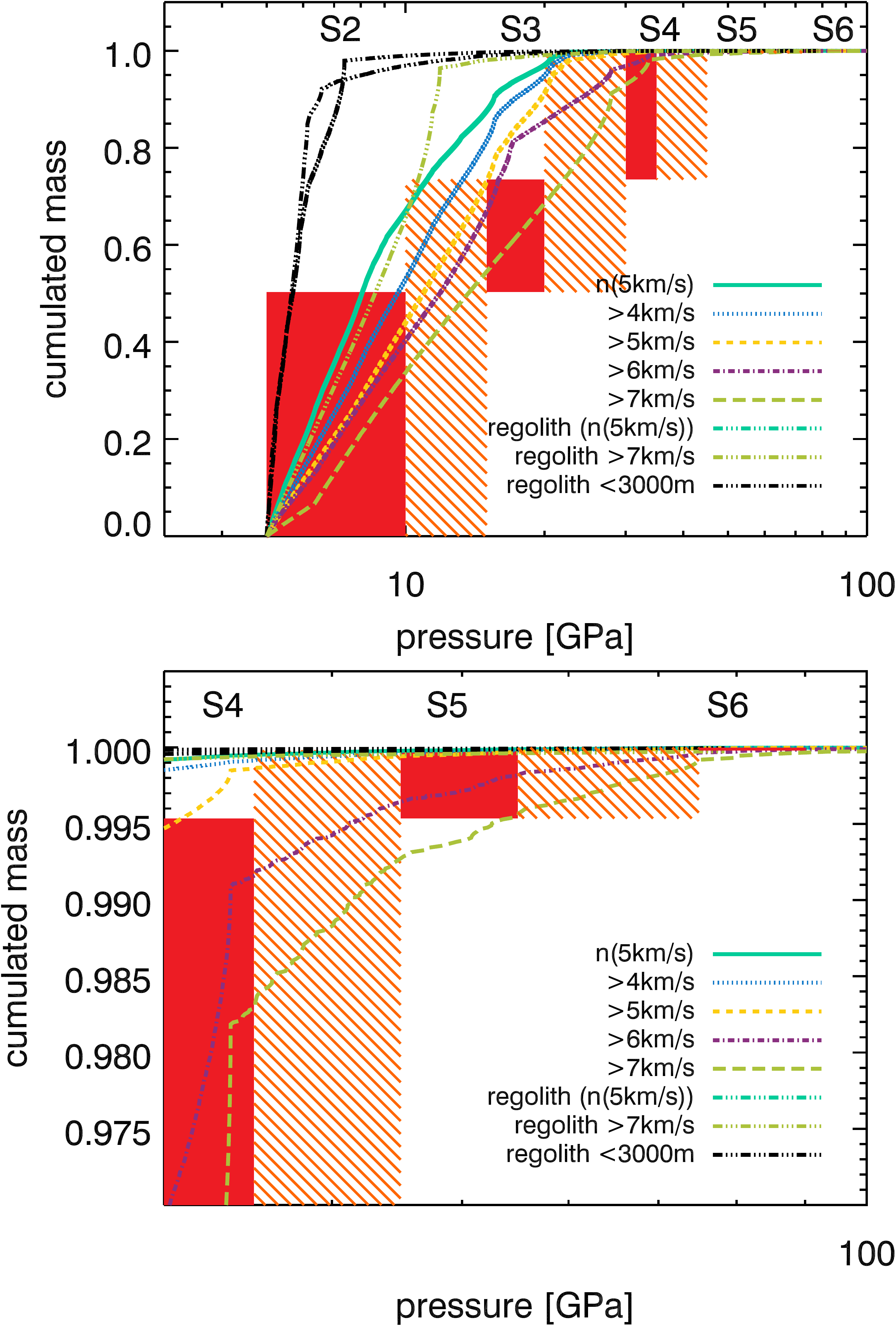}
		\caption{\label{fig:shock_0.600000_oclow_shock_ast_vel} Top: Comparison between the observed distribution of shock stages in ordinary chondrites with a petrologic type less than 4 and with a defined and unique shock stage (red boxes), and the maximum pressure experienced by the mass ejected in the last 20 Myrs in our simulations. The ordinate denotes the normalized cumulative mass of the meteorites and the excavated particles in our simulations. The dashed boxes refer to the transition zones between the shock stages, which we assign to both neighboring shock stages in terms of mass. The colored curves all refer to the F-LH-N simulation, but only surface elements that were hit by impacts above the stated velocity limit are cumulated to study the effect of the collision velocity. One can see that the best correlation is found for impacts above 4 $\mathrm{km\,s^{-1}}$.  Bottom: Zoom-in into shock stages 5 and 6. }
	\end{center}
\end{figure}

The distribution of shock stages of chondritic meteorites with low petrologic type in comparison with the numerical results of this study are shown in Fig. \ref{fig:shock_0.600000_oclow_shock_ast_vel}. On the ordinate, we plot the normalized cumulative mass of either the meteorites in the Meteoritical Bulletin Database or that of the excavated mass in our simulations during the last 20 Myrs, according to the typical CREA of the meteorites, which can be interpreted as a mean value of the real CREA ranging from 5 to 40 Myrs \citep{eugster2006irradiation}. For better visibility, the bottom panel of Fig. \ref{fig:shock_0.600000_oclow_shock_ast_vel} zooms into the highest shock stages S5 and S6. The dashed regions between the well-defined red shock-stage regions are transition zones, which we assigned to both neighboring shock stages. From Fig. \ref{fig:shock_0.600000_oclow_shock_ast_vel}, one can see that the cases with re-accretion of regolith do not fit the observed shock stages. The reason for this mismatch is that the overlaying km-thick regolith layer results in a much lower impact pressure, due to the considerably lower sound velocity in the granular regolith, according to Eq. \ref{eq:pressure_melosh_easy}. A better match between observed shock stages and simulated pressures can be found for those cases in which the impactors hit consolidated material. Here, a strong dependence of the pressure distribution function on the maximum impact velocity was found. The best agreement with all shock stages was found for those surface elements of the F-LH-N simulation that were hit at velocities higher than 6 $\mathrm{km\,s^{-1}}$, neglecting all surface elements that were not hit at all or at lower velocities. By comparing these results with those shown in Fig. \ref{fig:met_time_evoltuion}, we see that the filling factors for those cases without regolith are typically 0.98-0.99, whereas with regolith filling factors of 0.92 are achieved. These values are unfortunately so close together that the porosity values found in ordinary chondrites, namely $7.4 \pm 5.3$\% (corresponding to $\phi = 0.93 \pm 0.05$) for falls and $4.4 \pm 5.1$\% ($\phi = 0.96 \pm 0.05$) for finds, as reported by \citet{2008ChEG...68....1C}, can be explained by both cases. Here, it must also be taken into account that macroporosity (i.e., by cracks) could play a role in meteorites that cannot occur in our model.

We have shown above that asteroids with sizes in the 100 km range should retain most of the excavated material and, thus, should be covered with a many km thick regolith layer. However, this cannot explain the high shock stages found in ordinary chondrites, which, according to our simulations, can only be reproduced in a regolith-devoid case. We will discuss this discrepancy in the following section.

\subsection{An evolution model for asteroids and the formation of meteoroids}
In this subsection, we will propose an evolution model for asteroids and a formation model of meteoroids that is based on the results and discussion presented in the previous Sections.

The evolution of the asteroids starts after their formation 4.5 Gyrs ago as large and porous parent bodies. These bodies then are exposed to a steady bombardment of different sized impactors at random velocities during the time span until present. Most of the ejected mass in an impact is re-accreted by the $\gtrsim 100$ km-sized parent bodies and forms a layer, whose thickness depends on the size of the largest impactor. We predict the thickness of this layer through the size of the largest crater on the asteroid (see Fig. \ref{fig:regolith_thickness_crater_size}). \citet{schraepler2015} studied the packing density of a regolith layer under different gravity levels and found the volume filling factor to saturate at a value of 0.6 for thicknesses exceeding a few meters. This regolith layer shields the asteroid from a high degree of compaction on its surface. Nevertheless, a compacted crust is formed below the regolith layer (see Fig. \ref{fig:compaction_regolith}). A layer of regolith is also consistent with the formation of breccias that account for up to 15\% of all ordinary chondrites \citep{bischoff2006nature}. Material that escapes these parent bodies possesses time-independent properties (see Fig.  \ref{fig:met_time_evoltuion}). These meteoroids do not match the shock stages of real meteorites (see the regolith cases in Fig. \ref{fig:shock_0.600000_oclow_shock_ast_vel}).

The continuous bombardment leads to a significant change of the asteroid's shape by the cratering process. A comparison of our simulations to the surface shape of asteroid (21) Lutetia reveals that it can be explained best if the crater ejecta are gravitationally re-accreted (and, thus, form a regolith layer) and the largest impactor possesses a radius of 4 - 6 km (see Fig. \ref{fig:statistic}). With these parameters, a maximum crater diameter of about 40 - 60 km and a thickness of the regolith layer of $\sim 4$ km can be expected. This is in very good agreement with that largest crater found on (21) Lutetia.

From this, we conclude that chondritic meteorites cannot directly originate from parent bodies with a considerable regolith layer. Most ordinary chondrite groups exhibit a CREA of 5 - 40 Myrs, which is the time span the meteorites were exposed to the cosmic rays as bodies smaller than about a few meters in size, either as free-flying bodies or being on the surface of a larger body. Based on our above findings, we here propose that a very large impactor $\gtrsim 10$ Myrs ago lead to partial or full fragmentation of a large asteroid, thus exposing compacted material from the interior of the asteroid (see Fig. \ref{fig:compaction_regolith}). The impact occurred on the regolith and therefore did not produce high pressures of the surface material. The largest fragments of this impact were on the order of 1 km in size so that their lifetime against collisions in the asteroid belt was sufficiently long to deliver meteorites to the Earth. In a subsequent impact of a smaller body onto these regolith-free fragments, meteoroids were formed that experienced high impact pressures and led to meteorites with high shock stages, which is consistent with many ordinary chondrite falls from which there is evidence of multiple impacts at work during the full pathway to Earth \citep{llorca2005villalbeto}.
If the fragment is larger than 1 km in size, a significant amount of the ejecta would be re-accreted \citep{o1985impact}. As impacts at higher velocity also produce ejecta with higher velocity, which then would have an increased probability to escape the target body, this would explain the better fitting of the higher velocities to have produced the meteorites as visible in Fig. \ref{fig:shock_0.600000_oclow_shock_ast_vel}. From this point in time, the meteorites were being exposed to the cosmic rays for around $\sim 20$ Myrs.

Our scenario of large impacts playing a key role in the delivery of chondrites to Earth is also consistent with significant physical processes at work: excavation, fragmentation, brecciation and shock-induced annealing of these rocks. Just to cite some examples in our meteorite collections, genomic breccias are rocks containing clasts and matrix of the same compositional group, but of different metamorphic type that could be coming from different depths in the same parent asteroid (see, e.g., the review by \citet{bischoff2006nature}). These amazing rocks fit very well with mutual impacts under moderate shock pressures that can be roughly constrained due to the transformations induced in the rock-forming minerals. Another example are regolith, and fragmental breccias that evidence the compaction of surface materials by moderately large impacts. Sometimes the released energy by impacts is of such a magnitude that they produce shock melted rocks with unmelted clasts (known as impact melt breccias). Obviously, even although different collisional circumstances can occur in the real nature, the chondritic products found so far are reasonably well predicted by our model. We envision that these chondritic products were accumulated in the outer layer of collisionally processed asteroids, and can be also potentially transported by large impacts.

Carbonaceous chondrites are not the focus of this study, but they are in general agreement with the above evolution model. They show a typical CREA of only $\sim 10$ Myrs, which is lower compared to ordinary chondrites. As they need approximately the same time of $\sim 10$ Myrs to be transferred to Earth, the time span for the compaction after being ejected from their regolith-baring parent body is much shorter, and therefore a lower shock stage and volume filling factor can be expected. The relative rarity of shock members among carbonaceous chondrites in the meteorite collections can also be explained by the high bulk water contents of CI, CM and CR carbonaceous-chondrite groups, because they would be destroyed via explosive volatile expansion. In any case, volatile-poor groups like, e.g., CO and CV chondrites, have possibly not survived a high degrees of shock metamorphism. Probably they are less compacted by annealing and are too fragile to survive the long cascade of impact processes that deliver them to near-Earth space \citep{trigo2009tensile}.

\section{Conclusions}\label{sec:conclusions}

In this paper, we presented a model to explain the evolution of asteroids and the formation of meteoroids. The model follows the impact-compaction of 100-km sized asteroids since their formations as porous large spherical objects, due to the continuous bombardment of impactors between 0.1 m and 22 km in radius. From the size of the largest impact crater on the asteroid surface, the thickness of the regolith layer can be predicted. We found that the overall surface profile of the simulated asteroid to be roughly consistent with that of asteroid (21) Lutetia, which supports the assumption that asteroids were initially formed as large, almost spherical and porous objects.

We found that the chondrites originate most likely not from large, regolith-covered objects, but from smaller regolith-free asteroid fragments, whose sizes are such that their lifetime is larger than the CREA but which cannot retain considerable amounts of regolith.

\subsection*{Acknowledgements}
We thank the Deutsche Forschungsgemeinschaft (DFG) for support under Grant Bl 298/13-3 as part of the SPP 1385 ``The first 10 Millon Years of the Solar System''. We thank R. Di Sisto and G. de El\'{\i}a for stimulating discussions and G. de El\'{\i}a for providing us with his outputs of the present size-frequency distribution of bodies in the asteroid belt. M.G.P. thanks the Technische Universit\"at Braunschweig for support during her stay. J.B. and E.B. thank the visitors program of Facultad de Ciencias Astron\'omicas y Geof\'{\i}sicas, UNLP for financial support during their stay. This research was partially supported by Instituto Argentino de Radioastronom\'{\i}a(IAR), CCT-LA Plata, CONICET, Argentina and by the Argentine grant PIP 112-200901-00461 of CONICET. JMTR research was supported by the Spanish Ministry of Science and Innovation (project: AYA2011-26522). We thank Dennis Fr\"uhauff for 3d rendering support.



\bibliographystyle{aa}
\bibliography{literatur}

\end{document}